\documentclass[manuscript,screen]{acmart}

\usepackage{algpseudocode}
\usepackage{algorithm}
\usepackage{amsmath,amsfonts}
\usepackage{array}
\usepackage[caption=false,font=normalsize,labelfont=sf,textfont=sf]{subfig}
\usepackage{textcomp}
\usepackage{stfloats}
\usepackage{url}
\usepackage{verbatim}
\usepackage{graphicx}
\usepackage{zref}
\usepackage[normalem]{ulem}
\usepackage{tabularx}
\usepackage{setspace}
\usepackage{multirow}
\usepackage{booktabs}
\usepackage{listings}
\usepackage{xcolor}
 
\lstdefinelanguage{Java}{
    keywords={public, new, null, true, false, throws, assert, return}
}
\lstdefinelanguage{JSON}{
    keywords={true, false, null, yes, no},
    sensitive=true,
    morecomment=[l]{\#},
    morestring=[b]{"},
    keywordstyle=\bfseries\color{blue},
    commentstyle=\color{gray}
}
\lstdefinelanguage{YAML}{
    keywords={true, false, null, yes, no},
    sensitive=true,
    morecomment=[l]{\#},
    morestring=[b]{"},
    keywordstyle=\bfseries\color{blue},
    commentstyle=\color{gray},
    stringstyle=\color{green}
}
\usepackage[justification=centering]{caption}
\usepackage{framed}
\usepackage{tcolorbox}
\tcbuselibrary{skins}

\newcounter{DaveCommentCounter}
\newcommand{\ddu}[1]{
    \stepcounter{DaveCommentCounter}
    \textcolor{blue}{\textit{/**Dave's comment [\arabic{DaveCommentCounter}]: I don't understand the intended meaning in the next sentence. Please revise/delete/explain. **/}}
}

\newcommand{\dns}[1]{
    \stepcounter{DaveCommentCounter}
    \textcolor{blue}{\textit{/**Dave's comment [\arabic{DaveCommentCounter}]: I'm not sure that I have captured the intended meaning in the next sentence. Please check/confirm. **/}}
}

\newcounter{RubingCommentCounter}
\newcounter{SidaDengCounter}
\AtBeginDocument{%
  }

\setcopyright{acmlicensed}
\copyrightyear{2018}
\acmYear{2018}
\acmDOI{XXXXXXX.XXXXXXX}

\acmISBN{978-1-4503-XXXX-X/18/06}

\begin{document}

\title{LRASGen: LLM-based RESTful API Specification Generation}

\author{Sida Deng}
\email{3220002511@student.must.edu.mo}
\orcid{0009-0006-2819-3862}
\affiliation{
  \institution{School of Computer Science and Engineering, Macau University of Science and Technology}
  \city{Taipa}
  \state{Macau}
  \country{China}
  \postcode{999078}
}

\author{Rubing Huang}
\email{rbhuang@must.edu.mo}
\orcid{0000-0002-1769-6126}
\affiliation{
  \institution{School of Computer Science and Engineering, Macau University of Science and Technology}
  \city{Taipa}
  \state{Macau}
  \country{China}
  \postcode{999078}
}
\affiliation{
  \institution{Macau University of Science and Technology Zhuhai MUST Science and Technology Research Institute}
  \city{Zhuhai}
  \state{Guangdong Province}
  \country{China}
  \postcode{519099}
}

\author{Man Zhang}
\email{manzhang@buaa.edu.cn}
\orcid{0000-0003-1204-9322}
\affiliation{
  \institution{State Key Laboratory of Complex \& Critical Software Environment, Beihang University}
  \city{Beijing}
  \country{China}
  \postcode{100191}
}

\author{Chenhui Cui}
\email{3230002105@student.must.edu.mo}
\orcid{0009-0004-8746-316X}
\affiliation{
  \institution{School of Computer Science and Engineering, Macau University of Science and Technology}
  \city{Taipa}
  \state{Macau}
  \country{China}
  \postcode{999078}
}

\author{Dave Towey}
\email{dave.towey@nottingham.edu.cn}
\orcid{0000-0003-0877-4353}
\affiliation{
  \institution{School of Computer Science, University of Nottingham Ningbo China}
  \city{Ningbo}
  \state{Zhejiang}
  \country{China}
  \postcode{999078}
}

\author{Rongcun Wang}
\email{rcwang@cumt.edu.cn}
\orcid{0000-0002-9685-9893}
\affiliation{
  \institution{School of Computer Science and Technology, China University of Mining and Technology}
  \city{Xuzhou}
  \state{Jiangsu}
  \country{China}
  \postcode{221116}
}

\renewcommand{\shortauthors}{Deng et al.}

\begin{abstract}
    REpresentation State Transfer (REST) is an architectural style for designing web applications that enable scalable, stateless communication between clients and servers via common HTTP techniques. 
    Web APIs that employ the REST style are known as RESTful (or REST) APIs. 
    When using or testing a RESTful API, developers may need to employ its specification, which is often defined by open-source standards such as the OpenAPI Specification (OAS). 
    However, it can be very time-consuming and error-prone to write and update these specifications, which may negatively impact the use of RESTful APIs, especially when the software requirements change. 
    Many tools and methods have been proposed to solve this problem, such as Respector
    and Swagger Core.
    OAS generation can be regarded as a common text-generation task that creates a formal description of API endpoints derived from the source code.
    A potential solution for this may involve using Large Language Models (LLMs), which have strong capabilities in both code understanding and text generation.
    Motivated by this, we propose a novel approach for generating the OASs of RESTful APIs using LLMs: \textit{LLM-based RESTful API-Specification Generation} (LRASGen). 
    To the best of our knowledge, this is the first use of LLMs and API source code to generate OASs for RESTful APIs.
    Compared with existing tools and methods, LRASGen can generate the OASs, even when the implementation is incomplete (with partial code, and/or missing annotations/comments, etc.).
    To evaluate the LRASGen performance, we conducted a series of empirical studies on 20 real-world RESTful APIs:
    These studies examined the precision, recall, and F1-score when identifying the four main entities, i.e., endpoint methods and parameters, parameter constraints, and endpoint responses.
    The results show that two LLMs (GPT-4o mini and DeepSeek V3) can both support LARSGen to generate accurate specifications, 
    and LRASGen-generated specifications cover an average of 48.85\% more missed entities than the developer-provided specifications.
    In addition, compared with the state-of-the-art specification-generation techniques (Respector, Swagger Core, and Springdoc), we can observe: 
    1) For the 15 Java-based RESTful APIs, LRASGen outperforms all three state-of-the-art OAS generation techniques, which, on average, detect only
    75.67\% endpoint methods, 75.24\% endpoint parameters, 4.78\% parameter constraints, and 26.75\% endpoint responses.
    2) For the remaining five RESTful APIs (developed in Python and C\# programming languages), LRASGen achieves 100\% precision, recall, and F1-score for all four identification tasks; however,
    the other three techniques can not be successfully applied (because they are designed for Java-based APIs).
\end{abstract}

\ccsdesc[300]{Software and its engineering~Specification languages}
\ccsdesc[500]{Information systems~RESTful web services}

\keywords{RESTful API, OpenAPI Specification, Large Language Model}

\maketitle



\section{Introduction
\label{sec:1}}

REpresentational State Transfer (REST) \cite{fieldingArchitecturalStylesDesign2000} is an architectural style for designing web applications.
It uses typical World Wide Web infrastructure, such as the HTTP protocol, and is characterized by some key principles: 
Client-Server, Stateless, Cache, Uniform Interface, Layered System, and Code-On-Demand \cite{fieldingArchitecturalStylesDesign2000}.
Web APIs that follow the REST style can communicate and collaborate over the Internet, regardless of the programming language or framework used to develop them:
These kinds of APIs are called RESTful (or REST) APIs \cite{masse2011rest}.
RESTful APIs have been widely adopted, due to their simplicity and the ability to use existing web infrastructure. 
They typically use standard HTTP request methods (such as GET, POST, PUT, and DELETE) to perform create, retrieve, update, and delete operations on resources \cite{HTTPRequestMethods}.
The results of these requests are usually expressed using the standard HTTP response status codes: 
The \texttt{1xx} status code indicates information responses; 
\texttt{2xx} indicates successful responses; 
\texttt{3xx} indicates redirection messages; 
\texttt{4xx} indicates client-error responses; and
\texttt{5xx} indicates server-error responses \cite{HTTPResponseStatus}.
Modern service providers publish their services online in the form of RESTful APIs to allow subscribers to access text, audio, video, and other types of data \cite{neumannAnalysisPublicREST2021a}.

When the RESTful API goes online, it also publishes a specification describing various information about itself.
This specification can be in many forms, including the OpenAPI Specification (OAS) and the API blueprint:
Currently, OAS is the most popular \cite{TopAPISpecification}.
OAS~\cite{openapispecificationOpenAPISpecification}, formerly known as the Swagger Specification \cite{swaggerSwagger}, serves as the \textit{de facto} standard for describing RESTful APIs.
It enables both humans and computers to discover and understand the capabilities of a RESTful API, without requiring access to the source code, or necessitating network-traffic inspection. 
OAS serves as a cornerstone for designing, building, and testing RESTful APIs, providing a clear structure for developers to follow.
Despite the importance of the specifications, their quality may often be lacking, hindering effective API use and testing.
Poorly maintained or inaccurate specifications can lead to a disconnect between the RESTful API's intended use, and its actual implementation:
This can result in integration issues or potential security vulnerabilities.

As RESTful APIs evolve, keeping the specification and the implementation synchronized becomes increasingly difficult, potentially resulting in incomplete or inconsistent specifications.
A primary challenge relates to the manual creation and maintenance of the specification, which is time-consuming and error-prone.
This can be compounded by things like changing requirements, developers' poor writing skills, etc. 
To address these challenges, various tools and methodologies have been proposed to automate specification generation \cite{huangGeneratingRESTAPI2024, lercherGeneratingAccurateOpenAPI2024, swaggerSwagger, springdocopenapiSpringdocOpenAPI}.
However, these solutions often rely on annotations within the code, require manual intervention, or cannot identify complex API behaviors. 
Swagger Core \cite{swaggerSwagger} infers framework-specific and technique-specific annotations at runtime to generate specifications.
Similarly, Springdoc \cite{springdocopenapiSpringdocOpenAPI} also requires annotations for specification generation:
Without correct annotations, it is not possible to generate correct specifications.
These limitations have led to the exploration of static analysis techniques: 
Respector \cite{huangGeneratingRESTAPI2024} uses static analysis technology to generate the specifications directly from the source code, removing the need for additional annotations, and reducing the maintenance burden.
Whether it is an annotation-based approach, code-based methods, configuration-based techniques, or manual writing, understanding the source code is key to successfully generating an appropriate specification, such as OAS.

OAS generation can be regarded as a text-generation task that creates a formal description of API endpoints derived from the source code.
Recent studies \cite{houLargeLanguageModels2024} have shown that Large Language Models (LLMs) have excellent potential for code-understanding and text-generation tasks, which are precisely what is needed to understand API source code and generate OASs.
In this research, we introduce a novel approach: \textit{LLM-based RESTful API-Specification Generation} (LRASGen).
LRASGen is designed to revolutionize the way RESTful API specifications are generated.
LRASGen identifies endpoint methods, endpoint parameters, parameter constraints, and endpoint responses directly from the source code:
It uses LLMs to do this, ensuring that the generated specification is consistent with the API implementation, and automatically updates as the source code changes.
LLMs can be especially useful in situations where, for example, parts of the source code are not available, the structure is incomplete, or maybe some parameters are not fully defined:
LRASGen can then use the LLMs to generate reasonable specifications. 
To the best of our knowledge, LRASGen is the first technique to use an LLM and API source code to generate RESTful API specifications.

Taking the RESTful API source code as input, LRASGen goes through the following six steps to generate the specifications:
1) \textit{Identifying endpoint entry files:} 
The step involves analyzing the project structure, and locating the files that define the RESTful endpoints 
---
this can be done based on framework-specific criteria (such as annotations for Spring Boot, or configuration files for Django); 
2) \textit{Extracting and cleaning the endpoint code files:} 
This step ensures that only relevant code segments are retained, avoiding token limitations when interacting with the LLM; 
3) \textit{Identifying the endpoint methods:} 
Once the relevant files have been prepared, this step involves submitting these code files to the LLM (which uses designed prompts and examples to determine the endpoints); 
4) \textit{Identifying the endpoint parameters and responses:} 
This step requires a detailed decomposition of the API's endpoint parameters and responses, through the LLM;
5) \textit{Identifying the parameter constraints:} 
This step analyzes the source code and identifies parameter constraints for different types of endpoint parameters;
and 
6) \textit{Generating the OAS using the extracted information:} 
This step generates a standardized specification based on the OAS standard (v3.1.1) and the extracted endpoint information.
To explore the OAS-generation ability of LRASGen, we ran the LRASGen prototype on 20 real-world RESTful APIs, calculating the precision, recall, and F1-score for four tasks:
Identifying endpoint methods, endpoint parameters, parameter constraints, and endpoint responses.
Two different LLMs (GPT-4o mini and DeepSeek V3) are used.
The APIs are developed in three programming languages (Java, Python, and C\#) using six frameworks (Spring Boot, Jersey, Flask, Django, Web.py, and ASP.NET Core).
The experimental results show that both LLMs can support LARSGen to generate accurate specifications:
For the 15 Java-based RESTful APIs, LRASGen achieves 
99.46\% precision, 99.99\% recall, and 99.73\% F1-score for endpoint-method identification; and 
99.86\% precision, 98.86\% recall, and 99.33\% F1-score for endpoint-parameter identification. 
LRASGen also achieves 100\% precision, recall, and F1-score when identifying parameter constraints and endpoint responses.
LRASGen also identifies multiple entities missed by the developer-provided specifications: 
234 (out of 1308, 17.89\%) endpoint methods, 
2749 (out of 6023, 30.99\%) endpoint parameters, 
5530 (out of 96.30\%) parameter constraints, and 
1551 (out of 3087, 50.24\%) endpoint responses.
We also compare the performance of LRASGen with three state-of-the-art OAS generation techniques: 
Respector, Swagger Core, and Springdoc.
Based on this comparison, we can observe: 
1) For the 15 Java-based RESTful APIs, LRASGen outperforms all three state-of-the-art OAS generation techniques, which, on average, detect only
75.67\% endpoint methods, 75.24\% endpoint parameters, 4.78\% parameter constraints, and 26.75\% endpoint responses.

For the remaining five RESTful APIs (developed in Python and C\# programming languages), LRASGen achieves 100\% precision, recall, and F1-score for all four identification tasks;
the other three techniques can not be successfully applied (because they are designed for Java-based APIs). 

The main contributions of this paper are as follows:
\begin{enumerate}
    \item 
    We introduce the application of LLMs to automatically generate OpenAPI specifications directly from RESTful API source code.
    To the best of our knowledge, this is the first attempt to use GPT-4o mini and DeepSeek V3 for this task, establishing a novel and efficient approach for RESTful API-specification generation.
    \item 
    Based on previous research \cite{huangGeneratingRESTAPI2024}, we optimized the ground truth (GT) and proposed an enhanced ground truth (GT*) that used not only 15 Java RESTful APIs (from previous research \cite{huangGeneratingRESTAPI2024}), 
    but also an additional five real-world RESTful APIs developed using two other programming languages (Python and C\#), and three frameworks (Flask, Django, and Web.py).
    \item 
    Using LLM's outstanding abilities in natural language processing and code understanding, our approach supports the generation of RESTful APIs developed using various programming languages and frameworks.
    \item 
    We report on experiments conducted to compare the effectiveness of LRASGen with several state-of-the-art techniques for generating OpenAPI specifications:
    The results show that our approach generates more accurate and complete specifications.
    \item 
    We have implemented our approach as an open-source tool \cite{lrasgen} that generates OAS from the source code of a RESTful API.
\end{enumerate}

The rest of this paper is organized as follows.
Section \ref{sec:2} introduces the relevant background.
Section \ref{sec:3} presents an example to illustrate the problem that the LRASGen approach aims to overcome.
Section \ref{sec:4} explains the entire workflow of the LRASGen approach, and the specific details of each step.
Section \ref{sec:5} introduces the experimental setup, and Section \ref{sec:6} analyzes the experimental results, including examining the internal and external threats to validity.
Section \ref{sec:7} contains some further discussion and analyses. 
Section~\ref{sec:8} examines some related work.
Finally, Section~\ref{sec:9} concludes the paper, including some discussion of potential future work.

\section{Background
\label{sec:2}}

\subsection{RESTful API and OpenAPI Specification}

Introduced by Roy Fielding in 2000 \cite{fieldingArchitecturalStylesDesign2000}, REST is an architectural style for designing networked applications. 
It specifies a set of rules for creating APIs that allow resources to be accessed and modified over the network using HTTP.
REST emphasizes scalability of component interactions, generality of interfaces, and autonomous deployment of components (and intermediary components) to minimize interaction latency, enforce security, and encapsulate legacy systems.
It does not impose requirements on how it should be applied at the lower-level implementations (e.g., protocol stack layers or concrete middleware components), but does have several high-level design constraints, including the separation of client- and server-related concerns, statelessness, data caching, a consistent component interface, and code-on-demand \cite{fieldingArchitecturalStylesDesign2000}.

With REST, the implementation of the client and server can be done independently, without knowledge of the other:
Changes to client-side code do not affect the server’s operation, and vice versa.
This separation improves flexibility and scalability.
REST enables developers to interact with data stored on web servers using standard operations \cite{HTTPRequestMethods}.
REST-compliant systems are often called RESTful systems, with Web APIs that employ REST known as RESTful (or REST) APIs.
RESTful APIs typically have specifications that explain access to their services:
OAS is one of the most widely used standards for these specifications \cite{TopAPISpecification}.
However, writing and updating these specifications can be very time-consuming and error-prone:
If not done correctly, the use of the RESTful APIs may be impacted.
There is ongoing work into how to better test RESTful APIs \cite{arcuriAdvancedWhiteBoxHeuristics2023, arcuriAutomatedBlackWhiteBox2021, arcuriEnhancingSearchbasedTesting2022, atlidakisPythiaGrammarBasedFuzzing2020, atlidakisRestlerStatefulRest2019, corradiniAutomatedBlackboxTesting2023, corradiniRestatsTestCoverage2021, ehsanRESTfulAPITesting2022, zhangAdaptiveHypermutationSearchBased2022, wuCombinatorialTestingRESTful2022, viglianisiResttestgenAutomatedBlackbox2020, stallenbergImprovingTestCase2021, martin-lopezRESTestAutomatedBlackbox2021, kimAutomatedTestGeneration2022, kimEnhancingRESTAPI2023, kimAdaptiveRESTAPI2023}.

RESTful APIs can be built using various programming languages (e.g., Python, C\#, and Java) and frameworks (e.g., Spring Boot and Django).
Any type of data can be referred to as a resource in a RESTful API.
When a client request is made, a RESTful API responds 
by sending the caller a representation of the resource’s state.
Common data types that can be transferred over HTTP include plain text, HTML, and JSON (JavaScript Object Notation).
JSON \cite{JSON} is the most widely used file format because, contrary to its name, it is language-independent and understandable by both humans and machines \cite{WhichAPIData}. 

The OpenAPI Specification \cite{openapispecificationOpenAPISpecification}, previously known as the Swagger Specification \cite{swaggerSwagger}, defines a common language-independent interface to RESTful APIs.
It is a machine-readable interface definition language that serves as a standard for describing, producing, consuming, and visualizing web services.
It makes the service’s capabilities understandable to both humans and machines, without requiring access to network traffic analysis or the source code of the service.
When a remote service is specified correctly, users need minimal implementation logic to understand and utilize it.

These specifications enable automated code-generation tools to produce language-agnostic server stubs and client libraries, rigorous API testing frameworks to validate service behavior, and systematic development pipelines to leverage the structured interface definitions 
---
all derived directly from an OpenAPI Specification's machine-interpretable contract.
The specification simplifies RESTful API integration.
The OpenAPI Specification (OAS) formally defines an API specification as a structured object that can be serialized in either YAML or JSON format \cite{openapispecificationOpenAPISpecification}.
The primitive data types within this specification strictly adhere to those defined in the JSON Schema standard, enabling compatibility with JSON-based toolchains.

\subsection{Large Language Models}

Large Language Models (LLMs) have revolutionized the field of natural language processing (NLP), with an unprecedented ability to understand and generate human-like text.
These models, trained on vast corpora of text data, have shown remarkable capabilities in various NLP tasks, including language translation, sentiment analysis, and text summarization \cite{allenaiAI2WildBenchLeaderboard}.

The application of LLMs to code- and specification- generation is a particularly promising area of research. 
Early models, such as GPT (Generative Pre-trained Transformer), have been used to generate code snippets and assist in programming tasks.
LLMs have also been used to generate natural language descriptions of API endpoints and parameters \cite{kimEnhancingRESTAPI2023}.
These models can be trained on existing documentation to learn the nuances of API specification language, and to generate descriptions that are more accurate and readable. 
LLMs can also be used to fill in gaps in incomplete RESTful API specifications, providing a more comprehensive view of the API's functionality.

\section{A Motivating Example
\label{sec:3}}

RESTful APIs use HTTP request methods including GET, POST, PUT, and DELETE to define endpoints that facilitate create, retrieve, update, and delete operations on various resources.
Many enterprise-level RESTful APIs are Java-based; and Spring Boot \cite{springbootSpringBoot} and Jersey \cite{eclipsejerseyEclipseJersey} are the main frameworks used.
A significant number of high-quality RESTful APIs have been developed using diverse programming languages, demonstrating the versatility and widespread adoption of RESTful architecture:
The Poke API \cite{pokeapiPokeAPI} and Gramps Web API \cite{grampswebapiGrampsWebAPI}, for example, are well-documented examples implemented in Python; while the Bitwarden Server \cite{bitwardenserverBitwardenServer} was implemented in the C\# \cite{csharp} programming language.
These examples highlight the cross-language applicability of RESTful API development, and the importance of language-agnostic tools for API-specification generation.

Listing \ref{lst:Partial view of the source code of endpoint ``GET /statistics/projects'' in CatWatch API} shows a partial view of the ``GET /statistics/projects'' endpoint in the CatWatch API, using the Spring Boot framework.
The Spring Boot framework uses annotations to specify REST elements.
Lines 1-2 indicate that this is a GET method endpoint, and returns a collection of \texttt{ProjectStats} objects as a response.
Lines 3-11 indicate that the endpoint accepts optional query parameters, including \textit{organizations}, \textit{start\_date}, and \textit{end\_date}.
These parameters are used to filter the list of project statistics based on the specified organizations and date range.

\begin{lstlisting}[
float=t,
language=Java,
label={lst:Partial view of the source code of endpoint ``GET /statistics/projects'' in CatWatch API},
caption={Partial view of the source code of endpoint ``GET /statistics/projects'' in CatWatch API}
]
@RequestMapping(value = "/projects", method = RequestMethod.GET)
public ResponseEntity<Collection<ProjectStats>> statisticsProjectGet(
        @ApiParam(value = "List of github.com organizations to scan(comma seperated)", required = false)
        @RequestParam(value = Constants.API_REQUEST_PARAM_ORGANIZATIONS, required = false)
        String organizations,
        @ApiParam(value = "Date from which to start fetching statistics records from database(default = current date)")
        @RequestParam(value = Constants.API_REQUEST_PARAM_STARTDATE, required = false)
        String startDateString,
        @ApiParam(value = "Date till which statistics records will be fetched from database(default = current date)")
        @RequestParam(value = Constants.API_REQUEST_PARAM_ENDDATE, required = false)
        String endDateString
) throws java.text.ParseException {
    Date now = new Date();
    Date startDate = parseDate(startDateString, Date.from(now.toInstant().minus(30, ChronoUnit.DAYS)));
    Date endDate = parseDate(endDateString, now);
    List<Project> projects = null;
    if (organizations == null) {
        projects = projectRepository.findProjectsByDateRange(startDate, endDate);
    } else {
        Collection<String> orgs = StringParser.parseStringList(organizations, ",");
        projects = projectRepository.findProjectsByOrganizationNameAndDateRange(orgs, startDate, endDate);
    }
    assert (projects != null);
    List<ProjectStats> result = ProjectStats.buildStats(projects);
    // only top 10 by last score
    result.sort((ps1, ps2) -> -ps1.getScores().get(ps1.getScores().size() - 1)
        .compareTo(ps2.getScores().get(ps2.getScores().size() - 1)));
    ResponseEntity<Collection<ProjectStats>> res = new ResponseEntity<>(result.subList(0, 10), HttpStatus.\texttt{OK});
    return res;
}
\end{lstlisting}

When a RESTful API successfully processes a request and verifies the provided arguments, it returns a successful response status code (\texttt{200}); 
otherwise, it returns a response status code indicating that the request was malformed (\texttt{4xx}) or that a server error occurred (\texttt{5xx}).
Lines 13 to 14 in Listing \ref{lst:Partial view of the source code of endpoint ``GET /statistics/projects'' in CatWatch API} show that the endpoint method parses the input parameters and determines the date range.
If no \textit{start\_date} or \textit{end\_date} is provided, then the method defaults to the last 30 days and the current date, respectively.
Depending on whether or not the organization parameter is provided, the \texttt{ProjectStats} is fetched either for all organizations, or for the specified organizations, within the calculated date range.
The retrieved projects are then processed, their statistics computed, and the results sorted according to their latest score, in descending order.
Finally, it returns the top 10 \texttt{ProjectStats} with a response status code of \texttt{200}.
If the request parameters are invalid, or if the operation encounters an error, then the method throws appropriate exceptions and returns an error response with a \texttt{4xx} or \texttt{5xx} response status code.

According to the OpenAPI Specification standard, developers can declare multiple methods for a single path, with each endpoint uniquely identified by the combination of its path and HTTP request method.
Endpoint methods interact with resources through parameters embedded in the URL path/query or payload data transmitted in the request body.
The OpenAPI Specification provides specific keywords to enable the definition of the parameter attributes:
These include the name, location, data type, properties (e.g., format, default value, and examples), and constraints (e.g., minimum value, maximum length, and whether or not it is required).
The values of these parameters are restricted, to ensure that they result in valid HTTP responses.

Listing \ref{lst:Partial view of the LRASGen-generated OpenAPI Specification for endpoint ``GET /statistics/projects'' in CatWatch API} shows a partial specification of endpoint ``GET /statistics/projects'' in CatWatch API:
It begins by defining the endpoint path as ``/statistics/projects'', and specifies that it supports the GET method.
This endpoint is designed to retrieve statistical data related to projects.
Line 3 in Listing \ref{lst:Partial view of the LRASGen-generated OpenAPI Specification for endpoint ``GET /statistics/projects'' in CatWatch API} gets a summary with the JSON field \textit{``summary'': ``Fetch statistics of projects for given organizations''}, 
which concisely describes the functionality implemented by this endpoint.
The endpoint accepts three optional query parameters, which allow users to retrieve the filtered \texttt{ProjectStats}.
The parameters are defined within the \textit{parameters} field.
Each parameter is described with a name, location, optionality, description, and data type.
Taking the \textit{start\_date} parameter on Lines 7-14 as an example:
It is a string type, but also includes a format field specifying that it must follow the date-time format (YYYY-MM-DD`T'HH:mm:ssZ).
All three parameters are marked as optional (``\textit{required: false}'').
The response for a successful request is defined under the response status code \texttt{200}.
The description on line 19, the status code \texttt{200} indicates that the response is sent upon successful execution of the request.

\begin{lstlisting}[
float=t,
language=JSON,
label={lst:Partial view of the LRASGen-generated OpenAPI Specification for endpoint ``GET /statistics/projects'' in CatWatch API},
caption={Partial view of the LRASGen-generated OpenAPI Specification for endpoint ``GET /statistics/projects'' in CatWatch API}
]
"/statistics/projects": {
  "get": {
    "summary": "Fetch statistics of projects for given organizations",
    "operationId": "statisticsProjectGet",
    "parameters": [
      ...,
      {
        "name": "start_date",
        "in": "query",
        "required": false,
        "schema": {
          "type": "string",
          "format": "date-time",
          "description": "Date from which to start fetching statistics records from database (default = current date)"
        }
      },
      ...
    ],
    "responses": {
      "200": {
        "description": "An array of ProjectStats containing statistics of projects over the selected period of time.",
        "content": {
          "application/json": {
            "schema": {
              "$ref": "#/components/schemas/ProjectStats"
            }
          }
        }
      },
      ...
    }
  }
}
\end{lstlisting}

Based on a comparative analysis of the LRASGen-generated specification and the developer-provided specification of the CatWatch API, combined with an in-depth examination of the source code, we have identified the following issues:
1) The LRASGen-generated specification found a total of 14 endpoints, including six provided by the developer;
2) Of the six developer-provided endpoints, five were implemented with annotations, and the remaining one, ``GET /export'', was directly included, and contains three list objects: \texttt{List<Contributor>}, \texttt{List<Project>}, \texttt{List<Statistics>}.
and
3) Of the other eight endpoints not included in the developer-provided specification, four were marked as deprecated, and the other four were simply in the code, but are not in the developer-provided specification.

\section{The LRASGen Approach
\label{sec:4}}

\subsection{Framework
\label{sec:4.1}}

The proposed methodology to generate OpenAPI Specification using LLMs is structured into three parts: 
\textit{prompt preparation}, \textit{model calling}, and \textit{result processing}.

\begin{figure*}[!t]
    \centering
    \includegraphics[width=1.0\linewidth]{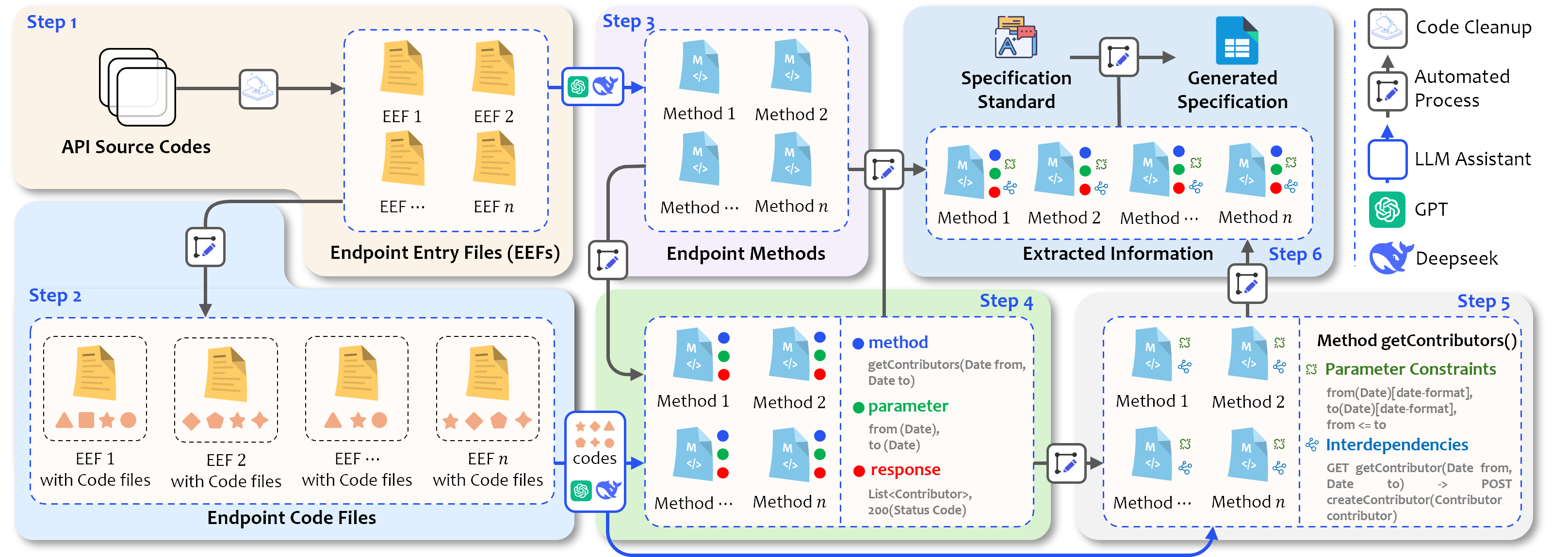}
    \Description[Overview of the LRASGen approach]{Overview of the LRASGen approach}
    \caption{Overview of the LRASGen approach}
    \label{FIG:Overview}
\end{figure*}

Due to LLM token limitations, the generation of OAS based on an entire project is not practical.
A better use of the LLM involves the extraction of the code files containing endpoint definitions from all the project files.
However, this must be done according to certain rules,
and the length of the context submitted to LLM should not be too long (to reduce the probability of truncation problems).

Excessive context length leads to information-forgetting and attention-distraction issues, so we split the source codes:
We start with a single endpoint, and extract its related code.
We then traverse the endpoint list and repeat the code extractions for each endpoint.
Finally, all the code related to all the endpoints is submitted to the LLM.
We also use zero/few-shot learning technologies \cite{romera-paredesEmbarrassinglySimpleApproach2015,wangGeneralizingFewExamples2021} to enhance LLMs' ability to accurately identify endpoint methods, endpoint parameters, parameter constraints, and endpoint responses.
This makes it possible for the LLM to return the fundamental data (endpoint method, parameter, etc) and sophisticated data (parameter constraints, variables, etc) that have been identified in the source code.

The entire specification-generation process takes the source code as input, and follows the latest version (v3.1.1) of the OpenAPI Specification standard.
Figure \ref{FIG:Overview} shows an overview of the LRASGen approach:
All the steps will be introduced in detail in the next section.
This systematic approach ensures the effective use of LLMs to produce high-quality specifications tailored to specific RESTful API implementations.

\subsection{Method
\label{sec:4.2}}

In this section, we present the LRASGen workflow, using the well-defined CatWatch API \cite{catwatchCatWatch} developed with the Spring Boot framework as a representative example. 

With this illustrative example, we clarify the step-by-step process of our approach, highlighting key mechanisms and operational aspects in practical scenarios.
LRASGen is a versatile approach that can be integrated into various programming languages and frameworks.
To demonstrate its applicability, we provide some practical examples of other programming languages and frameworks at different steps in the workflow.

\subsubsection{\textbf{Step 1: Identify Endpoint Entry Files}}
\label{sec:4.2.1}

As the first step of the entire workflow, it is essential to identify the endpoint source code.
Respector \cite{huangGeneratingRESTAPI2024} builds a framework-knowledge database by analyzing the Spring Boot \cite{springbootSpringBoot} and Jersey \cite{eclipsejerseyEclipseJersey} documents, and AutoOAS \cite{lercherGeneratingAccurateOpenAPI2024} obtains the endpoint source code by analyzing the project configuration files.
In this study, we selected several popular development frameworks (including Spring Boot \cite{springbootSpringBoot}, Jersey \cite{eclipsejerseyEclipseJersey}, Flask \cite{WelcomeFlaskFlask}, Django \cite{DjangoWebFrameworks}, Web.py \cite{web.pyWebpy}, and ASP.NET Core \cite{ASPNETCoreOpensource}).

The framework documentation was studied, and criteria for identifying endpoint entry files were established.
Listing \ref{lst:Partial view of Criteria for Identifying Endpoint Entry Files for Different Frameworks} is a partial view of the criteria for identifying endpoint entry files for different frameworks.
Lines 1 to 5 show the criteria of the Spring Boot, which is an annotation-based framework. 
For other configuration-based frameworks, such as Django \cite{DjangoWebFrameworks}, the criteria are shown in Lines 6 to 15.
LRASGen reads the configuration file defined in the \textit{``configuration\_files''} field and extracts the endpoint entry files in combination with the regex defined in the \textit{regex} field.

Algorithm \ref{alg:Identify Endpoint Entry Files} was designed to identify endpoint entry files by traversing a project directory and applying the specific criteria of the (annotation-based) framework:
Line 1 initializes an empty list, \textit{matched\_files}, and uses \textit{os.walk} to iterate through all the files listed in Line 2.
The files are then filtered by \textit{suffix}, and their contents are matched against predefined regular expressions.
The matching file paths are added to \textit{matched\_files}, which is returned as the final result, ensuring systematic identification of the relevant files.

\begin{algorithm}
\caption{Identify Endpoint Entry Files}
\label{alg:Identify Endpoint Entry Files}
\begin{algorithmic}[1]

\Require $directory$: Project root directory, $framework$: Framework configuration (suffix and regex patterns)
\Ensure $matched\_files$: List of files matching the framework criteria
\State $matched\_files \leftarrow []$
\ForAll{$(root, dirs, files)$ in $os.walk(directory)$}
    \ForAll{$file$ in $files$}
        \If{$file$ ends with $framework["suffix"]$} \Comment{suffix to filter the files}
            \State $file\_path \leftarrow os.path.join(root, file)$
            \State $content \leftarrow read(file\_path)$
            \If{any $regex$ in $framework["regex"]$ matches $content$} \Comment{check if this code file matched }
                \State $matched\_files.append(file\_path)$
            \EndIf
        \EndIf
    \EndFor
\EndFor
\State \Return $matched\_files$

\end{algorithmic}
\end{algorithm}

\begin{lstlisting}[
float=t,
language=JSON,
label={lst:Partial view of Criteria for Identifying Endpoint Entry Files for Different Frameworks},
caption={Partial view of Criteria for Identifying Endpoint Entry Files for Different Frameworks}
]
spring_boot = {
    "language": "java",
    "suffix": ".java",
    "regex": [ r"@(GetMapping|PostMapping|PutMapping|DeleteMapping|PatchMapping|RequestMapping|Controller|RestController)\([^)]*\)" ]
}
django = {
    "language": "python",
    "suffix": ".py",
    "regex": [
        r'urlpatterns\s*=\s*\[[^\]]*(path\([\'"]([^\'"]+)[\'"],\s*([^,]+)\)|re_path\([\'"]([^\'"]+)[\'"],\s*([^,]+)\))[^\]]*\]'
    ],
    "configuration_files": [
        "urls.py"
    ]
}
\end{lstlisting}

\subsubsection{\textbf{Step 2: Extract and Clean Endpoint Code Files}}
\label{sec:4.2.2}

After determining the endpoint entry file, it is necessary to start from the entry file and extract all the related code files in the project.
To avoid exceeding the token limitation, the extracted code files need to be cleaned up, and unnecessary code segments that are not related to the endpoint need to be removed.
Algorithm \ref{alg:Extract and Clean Endpoint Code Files} shows the whole process for the CatWatch API project.

\begin{algorithm}
\setstretch{0.96}
\caption{Extract and Clean Endpoint Code Files}
\label{alg:Extract and Clean Endpoint Code Files}
\begin{algorithmic}[1]

\Require $endpoint\_entry\_file\_paths$ is a list of valid Java file paths, and all paths are absolute
\Require $project\_root\_path$ is a valid project root directory path
\Ensure Returns a dictionary where keys are entry file paths and values are dictionaries containing cleaned code for the entry file and its related files
\State endpoint\_codes\_map $\gets$ empty dictionary
\For{each $entry\_file$ in $endpoint\_entry\_file\_paths$}
    \State $filtered\_files \gets$ \Call{filter\_files}{$entry\_file$, $project\_root\_path$}
    \State $all\_files \gets$ [$entry\_file$] + $filtered\_files$
    \State endpoint\_lines $\gets$ empty dictionary
    \For{each $file$ in $all\_files$}
        \State endpoint\_lines[$file$] $\gets$ \Call{read\_and\_clean\_codes}{$file$}
    \EndFor
    \State endpoint\_codes\_map[$entry\_file$] $\gets$ endpoint\_lines
\EndFor
\State \Return endpoint\_codes\_map

\Statex

\Function{filter\_files}{java\_file\_path, root\_path}
    \State class\_to\_file $\gets$ \Call{build\_class\_to\_file\_mapping}{$root\_path$}
    \State all\_imported\_files $\gets$ \Call{get\_all\_imported\_files}{$java\_file\_path$, $root\_path$, class\_to\_file}
    \State filtered\_files $\gets$ \Call{filter\_files\_by\_root}{$all\_imported\_files$, $root\_path$}
    \State \Return filtered\_files
\EndFunction

\Statex

\Function{build\_class\_to\_file\_mapping}{root\_path}
    \State class\_to\_file $\gets$ empty dictionary
    \For{each $file$ in \Call{recursively\_list\_files}{$root\_path$}} \Comment{recursively get all imported files}
        \If{$file$ ends with `.java`}
            \State content $\gets$ \Call{read\_file}{$file$}
            \State type\_name\_match $\gets$ \Call{find\_class\_or\_interface\_name}{content}
            \State package\_match $\gets$ \Call{find\_package\_name}{content}
            \If{type\_name\_match is valid}
                \State type\_name $\gets$ \Call{extract\_type\_name}{type\_name\_match}
                \State package\_name $\gets$ \Call{extract\_package\_name}{package\_match}
                \State class\_to\_file[type\_name] $\gets$ file
            \EndIf
        \EndIf
    \EndFor
    \State \Return class\_to\_file
\EndFunction

\Statex

\Function{get\_all\_imported\_files}{java\_file\_path, root\_path, class\_to\_file}
    \State file\_content $\gets$ \Call{read\_file}{$java\_file\_path$}
    \State imported\_classes $\gets$ \Call{find\_imported\_classes}{file\_content}
    \State imported\_files $\gets$ empty list
    \For{each $class\_name$ in imported\_classes}
        \If{$class\_name$ in class\_to\_file}
            \State imported\_files.append(class\_to\_file[$class\_name$])
        \EndIf
    \EndFor
    \State \Return imported\_files
\EndFunction

\end{algorithmic}
\end{algorithm}

\begin{algorithm}
\ContinuedFloat
\setstretch{0.96}
\caption{Extract and Clean Endpoint Code Files}
\begin{algorithmic}[1]

\Function{filter\_files\_by\_root}{files, root\_path}
    \State filtered $\gets$ empty list
    \For{each $file$ in $files$}
        \If{$file$ starts with $root\_path$}
            \State filtered.append($file$)
        \EndIf
    \EndFor
    \State \Return filtered
\EndFunction

\Statex

\Function{read\_and\_clean\_codes}{file}
    \State content $\gets$ \Call{read\_file}{$file$}
    \State cleaned\_content $\gets$ \Call{remove\_irrelevant\_lines}{content} \Comment{remove the code lines which not related to endpoint}
    \State cleaned\_content $\gets$ \Call{remove\_empty\_lines}{cleaned\_content}
    \State \Return cleaned\_content
\EndFunction

\end{algorithmic}
\end{algorithm}

\subsubsection{\textbf{Step 3: Identify Endpoint Methods}}
\label{sec:4.2.3}

Originally, we wanted to process the entire source code in a single LLM submission, to generate the specifications.
However, this proved unsatisfactory, with several critical issues arising, including excessive token usage, irrelevant responses, and instances of fabricated outcomes.
These challenges significantly hindered the effectiveness and accuracy of the automated specification generation.
We therefore refined our approach by segmenting the interaction process into multiple discrete steps.

To identify all the endpoint methods, we submit the endpoint entry file to the LLM and let it extract the endpoint information.
Listing \ref{lst:Partial view of Prompts and Response Format used in Step 3} shows a partial view of the prompts and response format used in this step.
For annotation-based frameworks that are defined in code files, only the endpoint entry file is submitted (for configuration-based frameworks, the configuration file is also submitted).
To avoid problems such as excessive token usage, other code files imported from the endpoint entry file are not included.

We submit an endpoint entry file as an attachment, and then use prompts to design a thinking chain to guide the LLM to identify the endpoints in the file.
We also provide some examples for the LLM to refer to, and provide the required return format (JSON).

\begin{lstlisting}[
float=t,
language=Python,
label={lst:Partial view of Prompts and Response Format used in Step 3},
caption={Partial view of Prompts and Response Format used in Step 3}
]
content = (
    f"Read the endpoint entry code (scoped from ## to ##): "
    f"##{endpoint_code_lines}##, and following these steps: "
    f"1.How many endpoints are included in the code? "
    f"2.For each endpoint, what is its HTTP_METHOD (e.g., GET, POST, ...)? "
    f"3.For each endpoint, what is its URL path? "
    f"4.For each endpoint, what is its method name? "
    f"At last, provide the result in JSON format, please strictly follow the example: {re_json_example}."
)

re_json_example = [{
    "endpoint_path": "/api/getUser",
    "http_method": "GET",
    "method_name": "getUser()"
}]
\end{lstlisting}

\subsubsection{\textbf{Step 4: Identify Endpoint Parameters and Responses}}
\label{sec:4.2.4}

Unlike Step 3 (Section \textit{\ref{sec:4.2.3}}), Step 4 requires a detailed decomposition of the API's endpoint methods, endpoint parameters, and endpoint responses, through the LLM.
This process involves distinct prompts and the provision of various examples to ensure precision and relevance in the generated outputs.
Listing \ref{lst:Partial view of Prompts and Response Format used in Step 4} shows a partial view of the prompts and response format used in this step.

An analysis of the framework documentation, supported by experienced developers' guidance, provides an understanding that endpoint-related information may be defined in multiple ways, in endpoint-entry files:
For example, with the Spring Boot framework, constant values can be referenced in annotations, with these values being defined in other files \cite{JavaBeanValidation}.
Lines 4, 7, and 10 in Listing \ref{lst:Partial view of the source code of endpoint ``GET /statistics/projects'' in CatWatch API} indicate that the \textit{value} attribute of \textit{@RequestParam} is imported from \textit{Constants.API\_REQUEST\_PARAM\_*}.

To ensure that no information is missed, the endpoint-entry file and other related files are submitted to LLM at the same time.
Similar to Step 3 (Section \textit{\ref{sec:4.2.3}}),  designed prompts, examples, and a specified response format are used to guide the LLM's analysis and return results.

\begin{lstlisting}[
float=t,
language=Python,
label={lst:Partial view of Prompts and Response Format used in Step 4},
caption={Partial view of Prompts and Response Format used in Step 4}
]
content = (
    f"Please read the endpoint codes (scoped from ## to ##): "
    f"##{endpoint_code_lines}##, and following these steps:"
    f"For the specific endpoint method named: {endpoint_method_name}, how many parameters are there? "
    f"1.For each parameter, what is its name, and type (e.g., string, number, integer, object, array, boolean)? Is it required? "
    f"2.What is this parameter located in (e.g., query or path)? "
    f"3.What is this parameter represent for? "
    f"4.What is this endpoint's response? how many responses do this endpoint returned?"
    f" what is the return HTTP status code for each one?"
    f" If an exception occurs, what is the exception message?"
    f" If any data is returned, what is the specific schema of the data?"
    f"At last, provide the result in JSON format, please strictly follow the example: {re_json_example}."
)

re_json_example = [{
    "endpoint_path": "/api/getUser",
    "endpoint_method": "GET",
    "description": "An Endpoint to Get User List",
    "parameters": [
        {
            "name": "str_param",
            "type": "string",
            "require": "true",
            "position": "query",
            "description": "some string parameter",
        }, {
            "name": "num_param",
            "type": "int",
            "require": "true",
            "position": "path",
            "description": "some number parameter",
        }, {
            "name": "bool_param",
            "type": "boolean",
            "require": "true",
            "position": "query",
            "description": "some boolean parameter",
        }
    ],
    "response": {
        "status_code": 200,
        "return_schema": [{"userName": "u", "password": "p", "birthday": "1970-01-01"},
                          {"userName": "u", "password": "p", "birthday": "1970-01-01"}],
        "exception": "NotFoundException"
    }
}]
\end{lstlisting}

\subsubsection{\textbf{Step 5: Identify Parameter Constraints}}
\label{sec:4.2.5}

Endpoint parameters are typically of three types: 
basic, user-defined, and mappings.
For basic types, different prompts and examples are used, according to the type.
For user-defined types, the LLM first locates the relevant code file, and then processes the fields in the user-defined type in the same way as for basic types.
For mappings, the LLM analyzes the entire endpoint code file, and obtains the parameter information from the code of the mapping read operation (e.g., \textit{Map.get()}, \textit{Map.values().get()}).

We designed several questions in the prompts to help the LLM to extract parameter-constraint information.
Some examples of such questions are:
``\textit{What is its type (e.g., string, number, integer, object, array, boolean)?}'',``\textit{Is this parameter required?}'', ``\textit{For a string-type parameter, what are its minimum length and maximum length?}'',``\textit{What is its default value?}''
We then designed a JSON schema with the information fields that the LLM was intended to extract.
Both the prompts and JSON schema are shown in Listing \ref{lst:Partial view of Prompts and Response Format used in Step 5}.

\begin{lstlisting}[
float=t,
language=Python,
label={lst:Partial view of Prompts and Response Format used in Step 5},
caption={Partial view of Prompts and Response Format used in Step 5}
]
content = (
    f"Please read the endpoint codes (scoped from ## to ##): "
    f"##{endpoint_code_lines}##, and following these steps:"
    f"For the specific parameter: {endpoint_parameter_name} in endpoint method named: {endpoint_method_name},"
    f"1.What is its type (e.g., string, number, integer, object, array, boolean)? Is it required? "
    f"2.For string type parameter, what is its minLength and maxLength? what is its default value?"
    f" If this string parameter is an enumeration, what is its enumeration?"
    f" If this string parameter is in date or pattern format, what is its date-time format or pattern?"
    f"3.For integer and number type parameter, what is its range? what is its default value? "
    f"4.For boolean type parameter, is it True or False? "
    f"5.For parameter in mapping, please read the endpoint codes and analyzed the above steps. "
    f"At last, provide the result in JSON format, please strictly follow the example: {re_json_example}."
)

re_json_example = [{
    "name": "str_param",
    "type": "string",
    "require": "true",
    "position": "query",
    "description": "some string parameter",
    "max_length": "128",
    "min_length": "16",
    "enum": ["enum1", "enum2", "enum3"],
    "format": "yyyy-mm-dd hh24:mi:ss",
    "default_value": "hello world"
}, {
    "name": "num_param",
    "type": "int",
    "require": "true",
    "position": "path",
    "description": "some number parameter",
    "min": 2,
    "max": 16,
    "default_value": 0
}, {
    "name": "bool_param",
    "type": "boolean",
    "require": "true",
    "position": "query",
    "description": "some boolean parameter",
    "default_value": True
}]
\end{lstlisting}

\subsubsection{\textbf{Step 6: Generate OpenAPI Specification using Extracted Information}}
\label{sec:4.2.6}

LRASGen first creates an empty JSON Object to be filled according to the given JSON schema \cite{JSONSchemaOpenAPI}.
This JSON object is the OAS to be generated.
In this JSON object, the \textit{paths} and \textit{components} fields are the two fields we mainly want to fill in.
The \textit{paths} field includes all the endpoints identified in Step 3 (Section \ref{sec:4.2.3}) and the endpoint-related information (such as HTTP method, request parameters, etc).
The \textit{components} field contains the structure of the custom parameters/data, which may be in the endpoint input or the endpoint response.
All designed fields and contents comply with the latest OpenAPI Specification standard.

\section{Experimental Methodology
\label{sec:5}}

In this section, we present the research questions (RQs) and the experimental design, which includes the subject APIs, LLM selection, evaluation metrics, and experiment procedure.

\subsection{Research Questions}
To systematically evaluate LRASGen, we used the following three RQs to address its accuracy, comprehensiveness, and comparative performance.

\textbf{RQ1: Can LRASGen generate an accurate specification?}

This research question assesses LRASGen's ability to generate accurate specifications, ensuring that they correctly reflect the API's behavior.

\textbf{RQ2: Do LRASGen-generated specifications cover behavior missed by developer-provided specifications?}

This research question investigates whether or not LRASGen can fill in the gaps in developer-provided specifications, providing a more complete behavioral representation.

\textbf{RQ3: How does LRASGen compare with the state-of-the-art RESTful API specification-generation techniques?}

This research question benchmarks LRASGen against state-of-the-art RESTful API specification-generation techniques, highlighting its relative strengths and limitations.

\subsection{Subject APIs}

In this study, we performed a comprehensive analysis on a set of 20 real-world RESTful APIs.
These APIs were carefully selected to represent a diverse range of technologies and frameworks.
We searched on GitHub using the following keyphrases: 
\textit{``Java RESTful API''}, 
\textit{``Python RESTful API''}, and
\textit{``C\# RESTful API''}.
We then selected APIs developed using Spring Boot, Jersey, Django, Flask, Web.py, and ASP.NET Core frameworks:
They all contained developer-provided specifications.
To ensure the quality and relevance of the selected projects, we applied the following two criteria for screening: 
1) The project must include a complete OpenAPI Specification provided by its developers; 
and 
2) The project must be actively maintained.
Based on these criteria, we narrowed down our initial collection of 27 to 20 RESTful APIs for analysis.
Table \ref{TAB:SubjectAPI} presents the detailed information of the 20 RESTful APIs used in this study, including each API's full and short name, framework, language, LoCs (number of lines of code), and the number of endpoints.
The first 15 APIs were developed in Java, and had previously been examined in other work \cite{huangGeneratingRESTAPI2024}.
The next four APIs were developed in Python, with Django as the framework for two projects, Flask for one, and Web.py for the last.
The last API was built using the ASP.NET Core framework, in the C\# programming language.

\begin{table*}[!t]
\footnotesize
    \centering
    \caption{20 Real-world RESTful APIs used in the study}
    \label{TAB:SubjectAPI}
    \setlength{\tabcolsep}{3mm}
    \begin{tabular}{lllllrr}
        \hline
        \textbf{No.} & \textbf{API}& \textbf{Short Name} & \textbf{Framework} & \textbf{Language} & \textbf{LoCs} & \textbf{\# Endpoints} \\
        \hline
        1& Digdag \cite{digdagDigdag}& Digdag& Jersey & Java& 54.8K & 41 \\
        2& enviroCar \cite{envirocarserverEnviroCarServer} & enviroCar & Jersey & Java& 22.7K & 128\\
        3& Features-Service \cite{featuresserviceFeaturesService}& Features-Service& Jersey & Java& 1.0K& 18 \\
        4& Gravitee.io \cite{graviteeapimanagementGraviteeAPIManagement} & Gravitee& Jersey & Java& 118.8K& 28 \\
        5& Kafka REST Proxy \cite{kafkarestproxyKafkaRestProxy}& Kafka & Jersey & Java& 19.4K & 69 \\
        6& Management API for Apache Cassandra \cite{managementapiforapachecassandraManagementAPIApache} & Cassandra & Jersey & Java& 24.4K & 50 \\
        7& RESTCountries \cite{restcountriesRestCountries} & RESTcountries & Jersey & Java& 88.2K & 27 \\
        8& Senzing \cite{senzingapiserverSenzingAPIServer} & Senzing & Jersey & Java& 30.9K & 34 \\
        9& CatWatch \cite{catwatchCatWatch}& CatWatch& Spring Boot& Java& 10.9K & 14 \\
        10 & CWA Verification Server \cite{cwaverificationserverCWAVerificationServer} & CWA & Spring Boot& Java& 45.4K & 5\\
        11 & OCVN \cite{ocvnOCVN}& OCVN& Spring Boot& Java& 9.0K& 278\\
        12 & Ohsome \cite{ohsomeapiOhsomeAPI}& Ohsome& Spring Boot& Java& 33.0K & 159\\
        13 & ProxyPrint \cite{proxyprintProxyprint}& ProxyPrint& Spring Boot& Java& 5.5K& 75 \\
        14 & Quartz Manager \cite{quartzmanagerQuartzManager}& Quartz& Spring Boot& Java& 2.3K& 15 \\
        15 & Ur-Codebin API \cite{ur-codebinapiUrCodebinAPI} & Ur-Codebin& Spring Boot& Java& 1.2K& 7\\
        16 & Poke API \cite{pokeapiPokeAPI}& Poke& Flask& Python& 43.6K & 97 \\
        17 & Gramps Web API \cite{grampswebapiGrampsWebAPI}& Gramps& Django & Python& 2.3K& 148\\
        18 & Jupyter Server \cite{jupyterserverJupyterServer}& Jupyter & Django & Python& 25.7K & 28 \\
        19 & Mlmmjadmin \cite{mlmmjadminMlmmjadmin}& Jupyter & Web.py & Python& 4.1K& 9\\
        20 & Bitwarden Server \cite{bitwardenserverBitwardenServer}& Bitwarden & ASP.net Core & C\# & 28.4K & 74 \\
        \hline
    \end{tabular}
\end{table*}

\subsection{LLM Selection}

The experiments used the commercial versions of the tools, provided by the supplier, hosted on their infrastructure.
This approach eliminated the need for local deployment, making available the tool's latest features, updates, and optimizations.
By using the cloud-based service, potential deployment complexities and version inconsistencies were avoided, ensuring a practical evaluation in the same context as its real-world use.
Furthermore, the commercial versions offer enhanced scalability, reliability, and security, which are essential for the validity and reproducibility of the results.
The experimental goal was to explore the efficacy of the LLMs in generating OpenAPI specifications, which are critical for defining and describing RESTful APIs \cite{openapispecificationOpenAPISpecification}.

Selection of the LLM was guided by indicators such as \textit{Info Seek} and \textit{Code} from the AI2 Wildbench Leaderboard \cite{allenaiAI2WildBenchLeaderboard} and the Deepseek official website \cite{deepseekDeepSeek}.
Other factors, such as consumption cost and response speed, were also considered.
Finally, GPT-4o mini and DeepSeek V3 were selected: 
Table \ref{TAB:LLMselection} presents their details.
GPT-4o mini \cite{openaiOpenAI} is a fast, affordable, small model for focused tasks:
It accepts both text and image inputs, and produces text outputs (including structured outputs).
It can be used for fine-tuning: 
Model outputs from a larger model (like GPT-4o) can be distilled into GPT-4o-mini to produce similar results at a lower cost and latency.
DeepSeek V3 \cite{deepseekDeepSeek} is an advanced language model developed by DeepSeek, known for its state-of-the-art performance, processing efficiency, and ability to comprehend and produce coherent and contextually-relevant text.
DeepSeek V3 has excellent accuracy and adaptability, and can be used for a wide range of applications, from natural language understanding \cite{guoDeepSeekCoderWhenLarge2024,gaoComparisonDeepSeekOther2025,manikChatGPTVsDeepSeek2025} to complex text-generation tasks \cite{guoDeepSeekCoderWhenLarge2024,gaoComparisonDeepSeekOther2025,manikChatGPTVsDeepSeek2025}.

\begin{table}[!t]
    \footnotesize
    \centering
    \caption{Two LLMs used in the study}
    \label{TAB:LLMselection}
    \begin{tabularx}
        {\linewidth}{p{2cm}ccX}
        \hline
        \textbf{Model} & \textbf{Context Window} & \textbf{Supplier} & \textbf{Description} \\
        \hline
        GPT-4o mini & 128K & OpenAI & {GPT-4o mini is a fast, affordable small model for focused tasks. It accepts both text and image inputs, and produces text outputs (including structured outputs).} \\
        \hline
        DeepSeek V3 & 64K & DeepSeek & {DeepSeek V3 is an advanced AI language model by DeepSeek, known for its high performance, efficiency, and ability to generate coherent, contextually precise text.} \\
        \hline
    \end{tabularx}
\end{table}

\subsection{Evaluation Metrics}
To evaluate the performance of the OpenAPI specification-generation technique, a ground truth (GT) was proposed in previous work \cite{huangGeneratingRESTAPI2024}.
Lercher et al.~\cite{lercherGeneratingAccurateOpenAPI2024} proposed an updated ground truth, GT+, which fixed some problems with GT (including correcting paths, adding missing prefixes, etc).
However, GT+ only selected seven RESTful APIs from GT, and only supported one programming language (Java)
and one framework (Spring Boot \cite{springbootSpringBoot}):
Because of this, GT+ may not be able to support the evaluation of LRASGen's performance with different programming languages and different frameworks. 
For this study, therefore, we created GT*, an enhanced version of GT (and GT+):
GT* not only covers all 15 Java RESTful APIs from GT (including various fixes), but also incorporates four Python RESTful APIs and one C\# API.
We have made GT* publicly available \cite{lrasgen}.

Each API in GT* was based on both its developer-provided specification and the source code.
We examined the RESTful API source code and the developer-provided specification:
We then identified the endpoint methods (such as URL path and HTTP method), their parameters (such as name, position, and data type), parameter constraints (such as minimum value, maximum length, and whether or not it is required), and endpoint responses (such as HTTP response status code and data).

After examination and identification, the ground truth, GT*, was established.
It was then compared with the LRASGen-generated specification:
{\em Precision} is defined as the ratio of correctly identified entities to the total entities identified in GT*; {\em recall} is defined as the ratio of correctly identified entities to the total number of correct entities in the GT* \cite{olson2008advanced}; and 
the {\em F1-score} is the harmonic mean of the precision and recall.
$TP$ (True Positive) is the number of identified entities that are in GT*;
$FP$ (False Positive) is the number of identified entities that are not in GT*; and 
$FN$ (False Negative) is the number of entities that were not identified, but that do exist in GT*:
The precision, recall, and F1-score can be calculated as follows:
\begin{equation}
    Precision = \frac{TP}{TP + FP}
\end{equation}

\begin{equation}
    Recall = \frac{TP}{TP + FN}
\end{equation}

\begin{equation}
    F1-score = 2 \times \frac{ Precision \times Recall}{Precision + Recall}
\end{equation}

All three metrics were calculated for:
(a) endpoint methods, 
(b) parameters for the identified endpoint methods, 
(c) constraints on the identified parameters, and 
(d) responses for the identified endpoint methods.
The accuracy of the LRASGen-generated specifications was also confirmed through cross-checks against the API source code.

\subsection{Experimental Procedure}

LRASGen automatically generated comprehensive specifications for the 20 real-world RESTful APIs, directly from their source code.
Table \ref{TAB:Time} presents the time consumption when running LRASGen using GPT-4o mini and DeepSeek V3, across these APIs.
The average running time of LRASGen, which was implemented entirely in Python, was 9 minutes and 36 seconds with GPT-4o mini and 6 minutes and 5 seconds with DeepSeek V3.
The shortest running time was under one minute (observed with the CWA API \cite{cwaverificationserverCWAVerificationServer}).

The accuracy of the LRASGen-generated specifications was evaluated against the enhanced ground truth (GT*). 
The generated specifications were also compared with developer-provided specifications to identify both the detected and the missed entities (endpoint methods, endpoint parameters, parameter constraints and endpoint responses).

Finally, to assess the effectiveness of LRASGen, 
we followed previous work \cite{huangGeneratingRESTAPI2024} and removed two techniques from the experiments:
Considering performance, adaptability, and other aspects, we removed AppMap \cite{appmapAppMap} (due to low accuracy, and its inability to recognize parameter and endpoint responses) and Springfox \cite{springfox} (due to poor adaptability, and it being last maintained at 2020).
Finally, we performed the comparative analysis against three state-of-the-art techniques: 
Respector, Swagger Core, and Springdoc.
This comparison highlighted the strengths and limitations of LRASGen in terms of the specification-generation accuracy, completeness, and efficiency.

\begin{table*}[!b]
    \footnotesize
    \centering
    \caption{Time Consumption of Running LRASGen Utilizing GPT-4o mini and
    DeepSeek V3 Across 20 real-world RESTful APIs}
    \label{TAB:Time}
    \setlength{\tabcolsep}{3.3mm}
    \begin{tabular}{lrr|rrr|rrr}
        \hline
        \multirow{2}*{\textbf{API}} & \multirow{2}*{\textbf{LoCs}} & \multirow{2}*{\textbf{\#Endpoint Methods}} & \multicolumn{3}{c}{\textbf{GPT-4o mini}} & \multicolumn{3}{c}{\textbf{DeepSeek V3}} \\
        \cline{4-9} &&& \textbf{min} & \textbf{max}& \textbf{avg} & \textbf{min}& \textbf{max} & \textbf{avg} \\
        \hline
        Digdag& 54.8K& 41 & 279& 386 & 332.5& 300 & 367& 333.5\\
        enviroCar & 22.7K& 128& 948& 1365 & 1156.5& 635 & 758& 696.5\\
        Features-Service& 1.0K & 18 & 223& 337 & 280& 414 & 418& 416\\
        Gravitee& 118.8K & 28 & 419& 526 & 472.5& 203 & 319& 261\\
        Kafka & 19.4K& 69 & 438& 690 & 564& 279 & 518& 398.5\\
        Cassandra & 24.4K& 50 & 1522& \textbf{3309}& 2415.5& 889 & \textbf{1128} & 1008.5\\
        RESTCountries & 88.2K& 27 & 160& 222 & 191& 148 & 197& 172.5\\
        Senzing & 30.9K& 34 & 245& 310 & 277.5& 167 & 349& 258\\
        CatWatch& 10.9K& 14 & 107& 153 & 130& 95& 172& 133.5\\
        CWA & 45.4K& 5& \textbf{39}& 67& 53.5 & \textbf{38} & 82 & 60 \\
        OCVN& 9.0K & 278& 1398& 1500 & 1,449& 665 & 734& 699.5\\
        Ohsome& 33.0K& 159& 2234& 2300 & 2267& 668 & 892& 780\\
        ProxyPrint& 5.5K & 75 & 870& 999 & 934.5& 754 & 887& 820.5\\
        Quartz& 2.3K & 15 & 73 & 100 & 86.5 & 50& 190& 120\\
        Ur-Codebin& 1.2K & 6& 40 & 48& 44 & 70& 210& 140\\
        Poke& 43.6K& 97 & 190& 256 & 223& 200 & 217& 208.5\\
        Gramps& 2.3K & 148& 75 & 135 & 105& 64& 177& 120.5\\
        Jupyter & 25.7K& 28 & 100& 129 & 114.5& 134 & 203& 168.5\\
        Mlmmj & 4.1K & 9& 275& 283 & 279& 222 & 368& 295\\
        Bitwarden & 28.4K& 74 & 127& 158 & 142.5& 119 & 289& 204\\
        \hline
        \multicolumn{3}{c|}{\textbf{\textit{Average}}} &488.15 & \textbf{663.65} & 575.90 &\textbf{305.70} & 423.75 & 364.73 \\
        \hline
    \end{tabular}
\end{table*}

\section{Experimental Results
\label{sec:6}}
In this section, we provide the experimental results to answer the three research questions.
Tables \ref{TAB:LLMresults1} and \ref{TAB:LLMresults2} present the evaluation results for RQ1:
LRASGen-generated specifications in the four identification tasks for 20 real-world RESTful APIs.
Table \ref{TAB:DeveloperResults} compares the amount of LRASGen-generated and developer-provided OASs across the four entities
(endpoint methods, endpoint parameters, parameter constraints, and endpoint response) (RQ2).
Tables \ref{TAB:ComparisonResults1} and \ref{TAB:ComparisonResults2} compare the performance of LRASGen with three state-of-the-art RESTful API specification-generation techniques (Respector, Swagger Core, and Springdoc).
Some symbols used in the tables are explained as follows:
\begin{itemize}
    \item 
    ``GT'' indicates the ground truth from previous work \cite{huangGeneratingRESTAPI2024}; 
    ``GT*'' indicates the enhanced ground truth proposed in this paper; 
    ``--'' indicates that ``GT'' was not applicable (Tables \ref{TAB:LLMresults1} and \ref{TAB:LLMresults2}), or that the API could not be run or that the tests failed (Tables \ref{TAB:ComparisonResults1} and \ref{TAB:ComparisonResults2});
    ``$\otimes$'' means these techniques could not generate any API specification (Tables \ref{TAB:ComparisonResults1} and \ref{TAB:ComparisonResults2});
    ``NA'' indicates that these techniques were not applicable (Tables \ref{TAB:ComparisonResults1} and \ref{TAB:ComparisonResults2}); and the values in brackets refer to the number of false positives in the LRASGen-generated specification.
   
    \item 
    In Table \ref{TAB:DeveloperResults}, the values in bold indicate that the OAS contains more than the other (for example, LRASGen-generated specification identified 41 endpoint methods in Digdag API while developer-provided specification had 39).
    
    \item 
    In Tables \ref{TAB:ComparisonResults1} and \ref{TAB:ComparisonResults2}, the values in bold indicate that the method achieved the best result in this task for this API.
\end{itemize}

\subsection{Answer to RQ1: Can LRASGen Generate An Accurate Specification?}


As shown in Tables \ref{TAB:LLMresults1} and \ref{TAB:LLMresults2},
LRASGen demonstrates a high level of precision, recall, and F1-score for the identification of endpoint methods, endpoint parameters, parameter constraints, and endpoint responses, because the precision, recall, and F1-score results for these four identification tasks score close to 100\%. The detailed observations can be described as follows:
\begin{itemize}
    \item \textit{\textbf{Endpoint Methods:}}
    LRASGen achieves an average of 99.46\% precision, 99.99\% recall, and 99.73\% F1-score for endpoint-method identification across the 15 Java-based APIs (Table \ref{TAB:LLMresults1}), identifying 952 (100.52\%) of the 947 endpoint methods. Among the 952 identified endpoint methods, seven are false positives (in three APIs: RESTCountries, OCVN, and ProxyPrint).
With the ``RESTCountries'' API, LRASGen identifies 28 out of 27 (GT*) endpoint methods:
The extra endpoint method is a false positive.
With the ``ProxyPrint'' API, LRASGen identifies 77 out of 75 (GT*) endpoint methods:
The two extra endpoint methods are false positives.
With the ``OCVN'' API, LRASGen identifies 280 out of 278 (GT*) endpoint methods:
Four of these are false positives, indicating that LRASGen actually identifies only 276 out of 278 endpoint methods.
Among the five remaining APIs (developed in Python and C\#), LRASGen achieves 100\% for precision, recall, and F1-score, identifying all 356 endpoint methods.

    \item \textit{\textbf{Endpoint Parameters:}}
LRASGen achieves an average of 99.86\% precision, 98.86\% recall, and 99.33\% F1-score for endpoint-parameter identification across the 15 Java-based APIs (Table \ref{TAB:LLMresults1}), identifying 8032 (97.68\%) of the 8222 endpoint parameters.
Among the identified 8032 endpoint parameters, LRASGen identifies 101 out of 99 in the ``Gravitee'' API:
The extra two endpoint parameters are false positives.
LRASGen failed to identify 291 endpoint parameters in four APIs.
This is because they are parameters of user-defined types:
The definition code snippets are not extracted and not added to the prompt, or the LLM ignored this part of the code snippet when interpreting the context.
Among the five remaining APIs (developed in Python and C\#), LRASGen achieves 100\% for both precision, recall, and F1-score, identifying all 840 endpoint parameters.

\item \textit{\textbf{Parameter Constraints:}} 
LRASGen achieves 100\% precision, recall, and F1-score when identifying the parameter constraints, finding all 5742 constraints across the 20 APIs.
LRASGen uses API source code to identify parameter constraints, which allows it to obtain parameter constraints not only from explicit definitions, but also from code snippets.

\item \textit{\textbf{Endpoint Responses:}} 
LRASGen identifies endpoint responses with 100\% precision, recall, and F1-score across the 20 APIs, detecting all 3087 responses.
Due to the inherent knowledge of the LLM, LRASGen not only recognizes the commonly used response status codes (such as \texttt{200}, \texttt{400}, and \texttt{500}), but also recognizes other user-defined codes that comply with the specification.
\end{itemize}

\begin{tcolorbox}[
    enhanced,
    width=\linewidth,
    colback=gray!10,
    colframe=gray!90,
    boxrule=1pt,
    arc=4pt,
    boxsep=5pt
]
\textbf{Summary:} LRASGen generates accurate specifications of 15 Java-based APIs with, on average, 
99.46\% precision, 99.99\% recall, and 99.73\% F1-score for endpoint methods;
99.86\% precision, 98.86\% recall, and 99.33\% F1-score for endpoint parameters;
100\% precision, 100\% recall, and 100\% F1-score for parameter constraints and endpoint responses.
For the other 4 Python and one C\# APIs, with all three metrics scoring 100\% on 4 identification tasks. (RQ1).
\end{tcolorbox}

\begin{table*}[!t] 
\footnotesize
    \centering
    \caption{Evaluating the accuracy of LRASGen-generated specifications in identifying the endpoint methods, endpoint parameters, parameter constraints, and endpoint responses for 15 Java RESTful APIs}
    \label{TAB:LLMresults1}
    \setlength{\tabcolsep}{1.4mm}
    \begin{tabular}{llrr|rrr|rrr|rrr}
        \hline
        \multirow{2}*{\textbf{Objective}} & \multirow{2}*{\textbf{API}} & \multirow{2}*{\textbf{GT}} & \multirow{2}*{\textbf{GT*}} & \multicolumn{3}{c|}{\textbf{GPT-4o mini}} & \multicolumn{3}{c|}{\textbf{DeepSeek V3}} & \multicolumn{3}{c}{\textbf{\textit{Average}}} \\
        \cline{5-13}& && & \textit{Precision}& \textit{Recall} & \textit{F1-score}& \textit{Precision} & \textit{Recall} & \textit{F1-score} & \textit{Precision} & \textit{Recall} & \textit{F1-score} \\
        \hline
        \multirow{15}*{Endpoint Methods}	&Digdag	&41	&41	&\textbf{1.00}	&\textbf{1.00}	&\textbf{1.00}	&\textbf{1.00}	&\textbf{1.00}	&\textbf{1.00}	&\textbf{1.00}	&\textbf{1.00}	&\textbf{1.00} \\	
	&enviroCar	&128	&128	&\textbf{1.00}	&\textbf{1.00}	&\textbf{1.00}	&\textbf{1.00}	&\textbf{1.00}	&\textbf{1.00}	&\textbf{1.00}	&\textbf{1.00}	&\textbf{1.00} \\	
	&Features-Service	&18	&18	&\textbf{1.00}	&\textbf{1.00}	&\textbf{1.00}	&\textbf{1.00}	&\textbf{1.00}	&\textbf{1.00}	&\textbf{1.00}	&\textbf{1.00}	&\textbf{1.00} \\	
	&Gravitee	&28	&28	&\textbf{1.00}	&\textbf{1.00}	&\textbf{1.00}	&\textbf{1.00}	&\textbf{1.00}	&\textbf{1.00}	&\textbf{1.00}	&\textbf{1.00}	&\textbf{1.00} \\	
	&Kafka	&74	&69	&\textbf{1.00}	&\textbf{1.00}	&\textbf{1.00}	&\textbf{1.00}	&\textbf{1.00}	&\textbf{1.00}	&\textbf{1.00}	&\textbf{1.00}	&\textbf{1.00} \\	
	&Cassandra	&50	&50	&\textbf{1.00}	&\textbf{1.00}	&\textbf{1.00}	&\textbf{1.00}	&\textbf{1.00}	&\textbf{1.00}	&\textbf{1.00}	&\textbf{1.00}	&\textbf{1.00} \\	
	&RESTCountries	&27	&27	&0.96	&\textbf{1.00}	&0.98	&0.96	&\textbf{1.00}	&0.98	&0.96	&\textbf{1.00}	&0.98 \\	
	&Senzing	&34	&34	&\textbf{1.00}	&\textbf{1.00}	&\textbf{1.00}	&\textbf{1.00}	&\textbf{1.00}	&\textbf{1.00}	&\textbf{1.00}	&\textbf{1.00}	&\textbf{1.00} \\	
	&CatWatch	&14	&14	&\textbf{1.00}	&\textbf{1.00}	&\textbf{1.00}	&\textbf{1.00}	&\textbf{1.00}	&\textbf{1.00}	&\textbf{1.00}	&\textbf{1.00}	&\textbf{1.00} \\	
	&CWA	&5	&5	&\textbf{1.00}	&\textbf{1.00}	&\textbf{1.00}	&\textbf{1.00}	&\textbf{1.00}	&\textbf{1.00}	&\textbf{1.00}	&\textbf{1.00}	&\textbf{1.00} \\	
	&OCVN	&278	&278	&\textbf{0.99}	&\textbf{0.99}	&\textbf{0.99}	&\textbf{0.99}	&\textbf{0.99}	&\textbf{0.99}	&\textbf{0.99}	&\textbf{0.99}	&\textbf{0.99} \\	
	&Ohsome	&159	&159	&\textbf{1.00}	&\textbf{1.00}	&\textbf{1.00}	&\textbf{1.00}	&\textbf{1.00}	&\textbf{1.00}	&\textbf{1.00}	&\textbf{1.00}	&\textbf{1.00} \\	
	&Quartz	&75	&75	&0.97	&\textbf{1.00}	&0.99	&0.97	&\textbf{1.00}	&0.99	&0.97	&\textbf{1.00}	&0.99 \\	
	&Proxyprint	&14	&15	&\textbf{1.00}	&\textbf{1.00}	&\textbf{1.00}	&\textbf{1.00}	&\textbf{1.00}	&\textbf{1.00}	&\textbf{1.00}	&\textbf{1.00}	&\textbf{1.00} \\	
	&Ur-Codebin	&7	&6	&\textbf{1.00}	&\textbf{1.00}	&\textbf{1.00}	&\textbf{1.00}	&\textbf{1.00}	&\textbf{1.00}	&\textbf{1.00}	&\textbf{1.00}	&\textbf{1.00} \\	\hline
\multirow{15}*{Endpoint Parameters}	&Digdag	&96	&97	&\textbf{1.00}	&\textbf{1.00}	&\textbf{1.00}	&\textbf{1.00}	&\textbf{1.00}	&\textbf{1.00}	&\textbf{1.00}	&\textbf{1.00}	&\textbf{1.00} \\	
        	&enviroCar	&351	&351	&\textbf{1.00}	&\textbf{1.00}	&\textbf{1.00}	&\textbf{1.00}	&\textbf{1.00}	&\textbf{1.00}	&\textbf{1.00}	&\textbf{1.00}	&\textbf{1.00} \\	
        	&Features-Service	&35	&35	&\textbf{1.00}	&\textbf{1.00}	&\textbf{1.00}	&\textbf{1.00}	&\textbf{1.00}	&\textbf{1.00}	&\textbf{1.00}	&\textbf{1.00}	&\textbf{1.00} \\	
        	&Gravitee	&99	&99	&0.98	&\textbf{1.00}	&0.99	&0.98	&\textbf{1.00}	&0.99	&0.98	&\textbf{1.00}	&0.99 \\	
        	&Kafka	&175	&175	&\textbf{1.00}	&\textbf{1.00}	&\textbf{1.00}	&\textbf{1.00}	&\textbf{1.00}	&\textbf{1.00}	&\textbf{1.00}	&\textbf{1.00}	&\textbf{1.00} \\	
        	&Cassandra	&94	&94	&\textbf{1.00}	&\textbf{1.00}	&\textbf{1.00}	&\textbf{1.00}	&\textbf{1.00}	&\textbf{1.00}	&\textbf{1.00}	&\textbf{1.00}	&\textbf{1.00} \\	
        	&RESTCountries	&35	&37	&\textbf{1.00}	&\textbf{1.00}	&\textbf{1.00}	&\textbf{1.00}	&\textbf{1.00}	&\textbf{1.00}	&\textbf{1.00}	&\textbf{1.00}	&\textbf{1.00} \\	
        	&Senzing	&156	&160	&\textbf{1.00}	&\textbf{1.00}	&\textbf{1.00}	&\textbf{1.00}	&\textbf{1.00}	&\textbf{1.00}	&\textbf{1.00}	&\textbf{1.00}	&\textbf{1.00} \\	
        	&CatWatch	&32	&32	&\textbf{1.00}	&\textbf{1.00}	&\textbf{1.00}	&\textbf{1.00}	&\textbf{1.00}	&\textbf{1.00}	&\textbf{1.00}	&\textbf{1.00}	&\textbf{1.00} \\	
        	&CWA	&14	&14	&\textbf{1.00}	&0.93	&0.96	&\textbf{1.00}	&0.93	&0.96	&\textbf{1.00}	&0.93	&0.96 \\	
        	&OCVN	&5002	&5002	&\textbf{1.00}	&0.98	&0.99	&\textbf{1.00}	&0.98	&0.99	&\textbf{1.00}	&0.98	&0.99 \\	
        	&Ohsome	&1937	&1937	&\textbf{1.00}	&0.96	&0.98	&\textbf{1.00}	&0.96	&0.98	&\textbf{1.00}	&0.96	&0.98 \\	
        	&Quartz	&150	&168	&\textbf{1.00}	&0.96	&0.98	&\textbf{1.00}	&0.96	&0.98	&\textbf{1.00}	&0.96	&0.98 \\	
        	&Proxyprint	&3	&7	&\textbf{1.00}	&\textbf{1.00}	&\textbf{1.00}	&\textbf{1.00}	&\textbf{1.00}	&\textbf{1.00}	&\textbf{1.00}	&\textbf{1.00}	&\textbf{1.00} \\	
        	&Ur-Codebin	&14	&14	&\textbf{1.00}	&\textbf{1.00}	&\textbf{1.00}	&\textbf{1.00}	&\textbf{1.00}	&\textbf{1.00}	&\textbf{1.00}	&\textbf{1.00}	&\textbf{1.00} \\	\hline
\multirow{15}*{Parameter Constraints}	&Digdag	&4	&112	&\textbf{1.00}	&\textbf{1.00}	&\textbf{1.00}	&\textbf{1.00}	&\textbf{1.00}	&\textbf{1.00}	&\textbf{1.00}	&\textbf{1.00}	&\textbf{1.00} \\	
	&enviroCar	&12	&71	&\textbf{1.00}	&\textbf{1.00}	&\textbf{1.00}	&\textbf{1.00}	&\textbf{1.00}	&\textbf{1.00}	&\textbf{1.00}	&\textbf{1.00}	&\textbf{1.00} \\	
	&Features-Service	&0	&37	&\textbf{1.00}	&\textbf{1.00}	&\textbf{1.00}	&\textbf{1.00}	&\textbf{1.00}	&\textbf{1.00}	&\textbf{1.00}	&\textbf{1.00}	&\textbf{1.00} \\	
	&Gravitee	&0	&59	&\textbf{1.00}	&\textbf{1.00}	&\textbf{1.00}	&\textbf{1.00}	&\textbf{1.00}	&\textbf{1.00}	&\textbf{1.00}	&\textbf{1.00}	&\textbf{1.00} \\	
	&Kafka	&0	&265	&\textbf{1.00}	&\textbf{1.00}	&\textbf{1.00}	&\textbf{1.00}	&\textbf{1.00}	&\textbf{1.00}	&\textbf{1.00}	&\textbf{1.00}	&\textbf{1.00} \\	
	&Cassandra	&3	&85	&\textbf{1.00}	&\textbf{1.00}	&\textbf{1.00}	&\textbf{1.00}	&\textbf{1.00}	&\textbf{1.00}	&\textbf{1.00}	&\textbf{1.00}	&\textbf{1.00} \\	
	&RESTCountries	&12	&24	&\textbf{1.00}	&\textbf{1.00}	&\textbf{1.00}	&\textbf{1.00}	&\textbf{1.00}	&\textbf{1.00}	&\textbf{1.00}	&\textbf{1.00}	&\textbf{1.00} \\	
	&Senzing	&17	&87	&\textbf{1.00}	&\textbf{1.00}	&\textbf{1.00}	&\textbf{1.00}	&\textbf{1.00}	&\textbf{1.00}	&\textbf{1.00}	&\textbf{1.00}	&\textbf{1.00} \\	
	&CatWatch	&3	&9	&\textbf{1.00}	&\textbf{1.00}	&\textbf{1.00}	&\textbf{1.00}	&\textbf{1.00}	&\textbf{1.00}	&\textbf{1.00}	&\textbf{1.00}	&\textbf{1.00} \\	
	&CWA	&0	&11	&\textbf{1.00}	&\textbf{1.00}	&\textbf{1.00}	&\textbf{1.00}	&\textbf{1.00}	&\textbf{1.00}	&\textbf{1.00}	&\textbf{1.00}	&\textbf{1.00} \\	
	&OCVN	&0	&4459	&\textbf{1.00}	&\textbf{1.00}	&\textbf{1.00}	&\textbf{1.00}	&\textbf{1.00}	&\textbf{1.00}	&\textbf{1.00}	&\textbf{1.00}	&\textbf{1.00} \\	
	&Ohsome	&58	&75	&\textbf{1.00}	&\textbf{1.00}	&\textbf{1.00}	&\textbf{1.00}	&\textbf{1.00}	&\textbf{1.00}	&\textbf{1.00}	&\textbf{1.00}	&\textbf{1.00} \\	
	&Quartz	&0	&228	&\textbf{1.00}	&\textbf{1.00}	&\textbf{1.00}	&\textbf{1.00}	&\textbf{1.00}	&\textbf{1.00}	&\textbf{1.00}	&\textbf{1.00}	&\textbf{1.00} \\	
	&Proxyprint	&0	&10	&\textbf{1.00}	&\textbf{1.00}	&\textbf{1.00}	&\textbf{1.00}	&\textbf{1.00}	&\textbf{1.00}	&\textbf{1.00}	&\textbf{1.00}	&\textbf{1.00} \\	
	&Ur-Codebin	&2	&14	&\textbf{1.00}	&\textbf{1.00}	&\textbf{1.00}	&\textbf{1.00}	&\textbf{1.00}	&\textbf{1.00}	&\textbf{1.00}	&\textbf{1.00}	&\textbf{1.00} \\	\hline
\multirow{15}*{Endpoint Responses}	&Digdag	&41	&85	&\textbf{1.00}	&\textbf{1.00}	&\textbf{1.00}	&\textbf{1.00}	&\textbf{1.00}	&\textbf{1.00}	&\textbf{1.00}	&\textbf{1.00}	&\textbf{1.00} \\	
	&enviroCar	&141	&201	&\textbf{1.00}	&\textbf{1.00}	&\textbf{1.00}	&\textbf{1.00}	&\textbf{1.00}	&\textbf{1.00}	&\textbf{1.00}	&\textbf{1.00}	&\textbf{1.00} \\	
	&Features-Service	&18	&35	&\textbf{1.00}	&\textbf{1.00}	&\textbf{1.00}	&\textbf{1.00}	&\textbf{1.00}	&\textbf{1.00}	&\textbf{1.00}	&\textbf{1.00}	&\textbf{1.00} \\	
	&Gravitee	&26	&57	&\textbf{1.00}	&\textbf{1.00}	&\textbf{1.00}	&\textbf{1.00}	&\textbf{1.00}	&\textbf{1.00}	&\textbf{1.00}	&\textbf{1.00}	&\textbf{1.00} \\	
	&Kafka	&108	&145	&\textbf{1.00}	&\textbf{1.00}	&\textbf{1.00}	&\textbf{1.00}	&\textbf{1.00}	&\textbf{1.00}	&\textbf{1.00}	&\textbf{1.00}	&\textbf{1.00} \\	
	&Cassandra	&60	&87	&\textbf{1.00}	&\textbf{1.00}	&\textbf{1.00}	&\textbf{1.00}	&\textbf{1.00}	&\textbf{1.00}	&\textbf{1.00}	&\textbf{1.00}	&\textbf{1.00} \\	
	&RESTCountries	&25	&72	&\textbf{1.00}	&\textbf{1.00}	&\textbf{1.00}	&\textbf{1.00}	&\textbf{1.00}	&\textbf{1.00}	&\textbf{1.00}	&\textbf{1.00}	&\textbf{1.00} \\	
	&Senzing	&34	&92	&\textbf{1.00}	&\textbf{1.00}	&\textbf{1.00}	&\textbf{1.00}	&\textbf{1.00}	&\textbf{1.00}	&\textbf{1.00}	&\textbf{1.00}	&\textbf{1.00} \\	
	&CatWatch	&10	&24	&\textbf{1.00}	&\textbf{1.00}	&\textbf{1.00}	&\textbf{1.00}	&\textbf{1.00}	&\textbf{1.00}	&\textbf{1.00}	&\textbf{1.00}	&\textbf{1.00} \\	
	&CWA	&7	&15	&\textbf{1.00}	&\textbf{1.00}	&\textbf{1.00}	&\textbf{1.00}	&\textbf{1.00}	&\textbf{1.00}	&\textbf{1.00}	&\textbf{1.00}	&\textbf{1.00} \\	
	&OCVN	&278	&664	&\textbf{1.00}	&\textbf{1.00}	&\textbf{1.00}	&\textbf{1.00}	&\textbf{1.00}	&\textbf{1.00}	&\textbf{1.00}	&\textbf{1.00}	&\textbf{1.00} \\	
	&Ohsome	&280	&449	&\textbf{1.00}	&\textbf{1.00}	&\textbf{1.00}	&\textbf{1.00}	&\textbf{1.00}	&\textbf{1.00}	&\textbf{1.00}	&\textbf{1.00}	&\textbf{1.00} \\	
	&Quartz	&75	&127	&\textbf{1.00}	&\textbf{1.00}	&\textbf{1.00}	&\textbf{1.00}	&\textbf{1.00}	&\textbf{1.00}	&\textbf{1.00}	&\textbf{1.00}	&\textbf{1.00} \\	
	&Proxyprint	&14	&22	&\textbf{1.00}	&\textbf{1.00}	&\textbf{1.00}	&\textbf{1.00}	&\textbf{1.00}	&\textbf{1.00}	&\textbf{1.00}	&\textbf{1.00}	&\textbf{1.00} \\	
	&Ur-Codebin	&7	&12	&\textbf{1.00}	&\textbf{1.00}	&\textbf{1.00}	&\textbf{1.00}	&\textbf{1.00}	&\textbf{1.00}	&\textbf{1.00}	&\textbf{1.00}	&\textbf{1.00} \\	\hline
    \end{tabular}
\end{table*}

\begin{table*}
    [!t] \footnotesize
    \centering
    \caption{Accuracy evaluation of LRASGen-generated specifications identifying the endpoint methods, endpoint parameters, parameter constraints, and endpoint responses for four Python RESTful APIs and a C\# RESTful API}
    \label{TAB:LLMresults2}
    \setlength{\tabcolsep}{1.6mm}
    \begin{tabular}{llrr|rrr|rrr|rrr}
        \hline
         \multirow{2}*{\textbf{Objective}}	&\multirow{2}*{\textbf{API}}	&\multirow{2}*{\textbf{GT}}	&\multirow{2}*{\textbf{GT*}}	&\multicolumn{3}{c|}{\textbf{GPT-4o mini}}	&\multicolumn{3}{c|}{\textbf{DeepSeek V3}}	&\multicolumn{3}{c}{\textbf{\textit{Average}}} \\								
        \cline{5-13}	&	&	&	&\textit{Precision}	&\textit{Recall}	&\textit{F1-score}	&\textit{Precision}	&\textit{Recall}	&\textit{F1-score}	&\textit{Precision}	&\textit{Recall}	&\textit{F1-score} \\		
        \hline														
\multirow{5}*{Endpoint Methods}	        &Poke	&--	&97	&\textbf{1.00}	&\textbf{1.00}	&\textbf{1.00}	&\textbf{1.00}	&\textbf{1.00}	&\textbf{1.00}	&\textbf{1.00}	&\textbf{1.00}	&\textbf{1.00} \\		
	        &Gramps	&--	&148	&\textbf{1.00}	&\textbf{1.00}	&\textbf{1.00}	&\textbf{1.00}	&\textbf{1.00}	&\textbf{1.00}	&\textbf{1.00}	&\textbf{1.00}	&\textbf{1.00} \\		
	        &Jupyter	&--	&28	&\textbf{1.00}	&\textbf{1.00}	&\textbf{1.00}	&\textbf{1.00}	&\textbf{1.00}	&\textbf{1.00}	&\textbf{1.00}	&\textbf{1.00}	&\textbf{1.00} \\		
	        &Mlmmj	&--	&9	&\textbf{1.00}	&\textbf{1.00}	&\textbf{1.00}	&\textbf{1.00}	&\textbf{1.00}	&\textbf{1.00}	&\textbf{1.00}	&\textbf{1.00}	&\textbf{1.00} \\		
	        &Bitwarden	&--	&74	&\textbf{1.00}	&\textbf{1.00}	&\textbf{1.00}	&\textbf{1.00}	&\textbf{1.00}	&\textbf{1.00}	&\textbf{1.00}	&\textbf{1.00}	&\textbf{1.00} \\		\hline
\multirow{5}*{Endpoint Parameters}	        &Poke	&--	&192	&\textbf{1.00}	&\textbf{1.00}	&\textbf{1.00}	&\textbf{1.00}	&\textbf{1.00}	&\textbf{1.00}	&\textbf{1.00}	&\textbf{1.00}	&\textbf{1.00} \\		
       	        &Gramps	&--	&488	&\textbf{1.00}	&\textbf{1.00}	&\textbf{1.00}	&\textbf{1.00}	&\textbf{1.00}	&\textbf{1.00}	&\textbf{1.00}	&\textbf{1.00}	&\textbf{1.00} \\		
       	        &Jupyter	&--	&20	&\textbf{1.00}	&\textbf{1.00}	&\textbf{1.00}	&\textbf{1.00}	&\textbf{1.00}	&\textbf{1.00}	&\textbf{1.00}	&\textbf{1.00}	&\textbf{1.00} \\		
       	        &Mlmmj	&--	&54	&\textbf{1.00}	&\textbf{1.00}	&\textbf{1.00}	&\textbf{1.00}	&\textbf{1.00}	&\textbf{1.00}	&\textbf{1.00}	&\textbf{1.00}	&\textbf{1.00} \\		
       	        &Bitwarden	&--	&86	&\textbf{1.00}	&\textbf{1.00}	&\textbf{1.00}	&\textbf{1.00}	&\textbf{1.00}	&\textbf{1.00}	&\textbf{1.00}	&\textbf{1.00}	&\textbf{1.00} \\		\hline
\multirow{5}*{Parameter Constraints}	        &Poke	&--	&49	&\textbf{1.00}	&\textbf{1.00}	&\textbf{1.00}	&\textbf{1.00}	&\textbf{1.00}	&\textbf{1.00}	&\textbf{1.00}	&\textbf{1.00}	&\textbf{1.00} \\		
       	        &Gramps	&--	&84	&\textbf{1.00}	&\textbf{1.00}	&\textbf{1.00}	&\textbf{1.00}	&\textbf{1.00}	&\textbf{1.00}	&\textbf{1.00}	&\textbf{1.00}	&\textbf{1.00} \\		
       	        &Jupyter	&--	&2	&\textbf{1.00}	&\textbf{1.00}	&\textbf{1.00}	&\textbf{1.00}	&\textbf{1.00}	&\textbf{1.00}	&\textbf{1.00}	&\textbf{1.00}	&\textbf{1.00} \\		
       	        &Mlmmj	&--	&1	&\textbf{1.00}	&\textbf{1.00}	&\textbf{1.00}	&\textbf{1.00}	&\textbf{1.00}	&\textbf{1.00}	&\textbf{1.00}	&\textbf{1.00}	&\textbf{1.00} \\		
       	        &Bitwarden	&--	&60	&\textbf{1.00}	&\textbf{1.00}	&\textbf{1.00}	&\textbf{1.00}	&\textbf{1.00}	&\textbf{1.00}	&\textbf{1.00}	&\textbf{1.00}	&\textbf{1.00} \\		\hline
\multirow{5}*{Endpoint Responses}	        &Poke	&--	&97	&\textbf{1.00}	&\textbf{1.00}	&\textbf{1.00}	&\textbf{1.00}	&\textbf{1.00}	&\textbf{1.00}	&\textbf{1.00}	&\textbf{1.00}	&\textbf{1.00} \\		
       	        &Gramps	&--	&591	&\textbf{1.00}	&\textbf{1.00}	&\textbf{1.00}	&\textbf{1.00}	&\textbf{1.00}	&\textbf{1.00}	&\textbf{1.00}	&\textbf{1.00}	&\textbf{1.00} \\		
       	        &Jupyter	&--	&56	&\textbf{1.00}	&\textbf{1.00}	&\textbf{1.00}	&\textbf{1.00}	&\textbf{1.00}	&\textbf{1.00}	&\textbf{1.00}	&\textbf{1.00}	&\textbf{1.00} \\		
       	        &Mlmmj	&--	&9	&\textbf{1.00}	&\textbf{1.00}	&\textbf{1.00}	&\textbf{1.00}	&\textbf{1.00}	&\textbf{1.00}	&\textbf{1.00}	&\textbf{1.00}	&\textbf{1.00} \\		
       	        &Bitwarden	&--	&247	&\textbf{1.00}	&\textbf{1.00}	&\textbf{1.00}	&\textbf{1.00}	&\textbf{1.00}	&\textbf{1.00}	&\textbf{1.00}	&\textbf{1.00}	&\textbf{1.00} \\		
        \hline
    \end{tabular}
\end{table*}

\subsection{Answer to RQ2: Do LRASGen-Generated Specifications Cover Behavior Missed by Developer-Provided Specifications?}
Based on Table \ref{TAB:DeveloperResults}, we have the following observations:
LRASGen-generated specifications often provide more comprehensive coverage of RESTful API behavior than the developer-provided ones.
This is mainly because LRASGen extracts information directly from the source code, ensuring that the specification aligns with the actual implementation.
In contrast, developer-provided specifications may be incomplete or outdated, due to human error or the natural evolution of the software.
LRASGen identifies multiple entities missing from the developer-provided specifications: 
234 (out of 1308, 17.89\%) endpoint methods, 
2749 (out of 6023, 30.99\%) endpoint parameters, 
5530 (out of 96.30\%) parameter constraints, and 
1551 (out of 3087, 50.24\%) endpoint responses.

\begin{itemize}
    \item \textit{\textbf{Endpoint Methods:}}
One of the key areas where LRASGen outperforms developer-provided specifications is in the identification of additional endpoints: 
With the CatWatch API, for example,
LRASGen identifies seven endpoints that are not documented in the developer-provided specification.
Among these, four are marked as deprecated in the source code, which may explain why they are omitted from the developer-provided specification.

    \item \textit{\textbf{Endpoint Parameters:}}
LRASGen identifies 2749 endpoint parameters missing from the developer-provided specifications. 
An analysis of these parameters reveals that they are usually encapsulated in the request body, or in deeply nested user-defined types.

    \item \textit{\textbf{Parameter Constraints:}}
LRASGen also often identifies parameter constraints that are not explicitly documented:
With the RESTCountries API, for example, LRASGen correctly identifies constraints (such as date formats for query parameters) that are not been included in the developer-provided specification.
This is because LRASGen analyzes the source code to infer constraints, even when they are not explicitly stated.
The developer-provided specifications for most of the Java RESTful APIs, for example, rarely contain constraints:
Maybe the developer does not record them when writing the specification, or maybe the tool does not identify them when generating the specification. 
These constraints
---
like \textit{required}, \textit{minValue}, \textit{maxValue}, \textit{date-time format} and \textit{enumeration}
---
do exist, and will restrict the input-parameter values of the endpoint.
For example, in Ur-Codebin API, the parameter \textit{paste\_syntax} in query position of endpoint ``\textit{GET /api/paste/public}'' is an enumeration type, and it must take values from the enumeration list (i.e., ``\textit{NONE}'', ``\textit{JAVA}'', ``\textit{CSHARP}'', ``\textit{Python}'', ``\textit{JAVASCRIPT}'', ``\textit{GO}'', ``\textit{CLANG}'', ``\textit{CPLUSPLUS}'', ``\textit{PHP}'', ``\textit{SWIFT}'', ``\textit{LUA}'', ``\textit{RUBY}'', ``\textit{MYSQL}'', and ``\textit{POSTGRESQL}'').

    \item \textit{\textbf{Endpoint Responses:}}
LRASGen extracts the response by analyzing the processing logic of each branch in the source code, and the final results of each branch.
This ensures that the LRASGen-generated specification includes, 
for example, the endpoint \textit{``POST /api/account/signup''} in the Ur-Codebin API has four possible responses: 
1) \texttt{200} response status code, with the description ``\textit{Successfully signed up for a new account.}'', and a returned data; 
2) \texttt{400} response status code, with the description ``\textit{Email is not in the correct format. Make sure the email is valid.}''; 
3) \texttt{409} response status code, with the description ``\textit{Username provided is already in use. Please use another username.}''; 
and 
4) \texttt{409} response status code, with the description ``\textit{Email provided is already in use. Please use another email.}''.
These four responses correspond to different instructions to guide users or downstream developers on how to use the API.
\end{itemize}

The comprehensive coverage provided by LRASGen is also due to its ability to handle incomplete or partially-provided code.
Unlike manual documentation, which relies on developers to update the specification as the code evolves, LRASGen can generate specifications even when some parts of the code are missing:
With the Ur-Codebin API, for example, LRASGen successfully generates a specification despite there being some incomplete code segments.
This flexibility is essential for maintaining up-to-date documentation in fast-paced development environments.

\begin{tcolorbox}[
    enhanced,
    width=\linewidth,
    colback=gray!10,
    colframe=gray!90,
    boxrule=1pt,
    arc=4pt,
    boxsep=5pt
]
\textbf{Summary:}
LRASGen identifies multiple entities missing from the developer-provided specifications: 
234 (out of 1308, 17.89\%) endpoint methods, 
2749 (out of 6023, 30.99\%) endpoint parameters, 
5530 (out of 96.30\%) parameter constraints, and 
1551 (out of 3087, 50.24\%) endpoint responses.
\end{tcolorbox}

\begin{table*}
    [!t] \footnotesize
    \centering
    \caption{Comparing LRASGen-generated with developer-provided OpenAPI
    Specifications for 20 real-world RESTful APIs}
    \label{TAB:DeveloperResults}
    \setlength{\tabcolsep}{2.9mm}
    \begin{tabular}{l|rr|rr|rr|rr}
        \hline
        \multirow{2}*{\textbf{API}} & \multicolumn{2}{c|}{\textbf{Endpoint Methods}} & \multicolumn{2}{c|}{\textbf{Endpoint Parameters}} & \multicolumn{2}{c|}{\textbf{Parameter Constraints}} & \multicolumn{2}{c}{\textbf{Endpoint Responses}} \\
        \cline{2-9} & Developer & LRASGen &Developer & LRASGen& Developer & LRASGen &Developer & LRASGen \\
        \hline
        Digdag& 39& \textbf{41} & 72& \textbf{97}& 0&\textbf{112}& 36 &\textbf{85}\\
        enviroCar & 77& \textbf{128}& 84& \textbf{351}& 0&\textbf{71}& 44 &\textbf{201}\\
        Features-Service& 18& 18 & 35&35 & 0&\textbf{37}& 7&\textbf{35}\\
        Gravitee& 26& \textbf{28} & 99&101 (2) & 0&\textbf{59}& 21 &\textbf{57}\\
        Kafka & 41& \textbf{69}& 101 &\textbf{175} & 0&\textbf{265}& 0&\textbf{145}\\
        Cassandra & 50& 50 & 94& 94& 0&85& 25 &87\\
        RESTCountries & 10& \textbf{28} (1)& 20&\textbf{37} & \textbf{0}&24& 0&\textbf{72}\\
        Senzing & 34& 34 & 150 &\textbf{160} & 13 &\textbf{87}& 34 &\textbf{92}\\
        CatWatch& 6 & \textbf{14} & 28&\textbf{32} & 1&\textbf{9}& 6&\textbf{24}\\
        CWA & 5 & 5& 5 & \textbf{13}& 0&11& 3&15\\
        OCVN& 190 & \textbf{280 (4)} & 2834 &\textbf{4888} & 0&\textbf{4459}& 160&\textbf{664}\\
        Ohsome& 135 & \textbf{159}& 1608 &\textbf{1867} & 0&\textbf{75}& 123&\textbf{449}\\
        ProxyPrint& 71& \textbf{77 (2)} & 44&\textbf{161} & 0&\textbf{228}& 67 & \textbf{127}\\
        Quartz& 10& 15 & 3 &\textbf{7} & 0&\textbf{10}& 6& \textbf{22}\\
        Ur-Codebin& \textbf{6} & \textbf{6}& 6 &\textbf{14} & 2&\textbf{14}& 4& \textbf{12}\\
        Poke& 97& 97 & 192 &192 & 49 &49& 97 & 97\\
        Gramps& 148 & 148 & 488 &488 & 84 &84& 591& 591\\
        Jupyter & 28& 28 & 20&20 & 2&2& 56 & 56\\
        Mlmmj & 9 & 9 & 54&54 & 1&1& 9& 9\\
        Bitwarden & 74& 74 & 86&86 & 60 &60& 247& 247\\
        \hline
        \textbf{\textit{Sum}} & 1074& \textbf{1308 (7)} & 6023&\textbf{8872 (2)} & 212 &\textbf{5742}& 1536& \textbf{3087}\\
        \hline
    \end{tabular}
\end{table*}

\subsection{Answer to RQ3: How Does LRASGen Compare with Alternative State-Of-The-Art RESTful API-Specification Generation Techniques?}
As shown in Tables \ref{TAB:ComparisonResults1} and \ref{TAB:ComparisonResults2}, 
LRASGen can process APIs developed in multiple programming languages (including Java, Python, and C\#) and frameworks (including Spring Boot, Django), whereas traditional tools (such as Swagger Core and Springdoc) are limited to specific frameworks (e.g., Spring Boot), which restricts their applicability: Swagger Core and Springdoc, for example, are unable to generate specifications for Python-based APIs (like Poke API and Gramps Web API), whereas LRASGen successfully processes them.
This demonstrates LRASGen's ability to generalize across different programming languages, making it a more versatile tool for real-world applications. The detailed observations can be categorized as follows:
\begin{itemize}
    \item \textit{\textbf{Endpoint Methods:}}
    In total, LRASGen identifies 945 (out of 952, 99.26\%) endpoint methods while Respector identifies 946 (99.36\%), for the 15 Java-based APIs.
    In contrast, Swagger Core identifies 40 (out of 45, 88.89\%) endpoint methods on the two Jersey APIs, and Springdoc find 173 (out of 179, 96.64\%) endpoint methods on the five Spring Boot APIs.
    Respector, while achieving high precision, also misses some endpoints: 
    This may have been because Respector relies on static analysis, which can be less effective at handling complex code structures \cite{huangGeneratingRESTAPI2024}.
    LRASGen, in contrast, uses LLMs to analyze the code, allowing it to identify endpoints even when they are not explicitly annotated.
    
    \item \textit{\textbf{Endpoint Parameters:}}
    LRASGen and Respector are applicable on all 15 Java-based APIs, Swagger Core only supports Jersey APIs, and Springdoc only supports Spring Boot APIs, 
    the ability of LRASGen to identify endpoint parameters is slightly better than the other three techniques:
    In total, LRASGen identifies 8032 (out of 8222, 97.68\%) endpoint parameters while Respector identifies 7977 (97.02\%), with the 15 Java-based APIs.
    In contrast, Swagger Core identifies 69 (out of 72, 95.83\%) endpoint parameters on the two Jersey APIs, and Springdoc identifies 1936 (out of 1983, 97.62\%) on the five Spring Boot APIs.
    
    \item \textit{\textbf{Parameter Constraints:}}
    LRASGen is the only tool that is able to consistently identify the parameter constraints:
    In total, LRASGen identifies all 5546 parameter constraints, while Respector only identifies 31 (0.56\%).
    Swagger Core and Springdoc could hardly identify any parameter constraints (two of 61, 3.3\%, for Swagger Core, and one of 98, 1.02\%, for Springdoc).
    With the RESTCountries API, for example, LRASGen correctly identifies parameter constraints such as date formats for query parameters, which are not identified by Swagger Core or Springdoc.
    Respector only identifies a few constraints, possibly because it relies on static analysis (which is less effective at inferring constraints from the code).
    
    \item \textit{\textbf{Endpoint Responses:}}
    LRASGen is better at identifying endpoint responses, particularly in RESTful APIs with complex response structures.
    With the Ohsome API, for example, LRASGen identifies 280 responses, compared to 261 by both Respector and Swagger Core.
    This is because LRASGen analyzes the response logic in the source code, directly extracting detailed information:
    LRASGen identifies all 2087 endpoint responses, 
    Respector identifies 1055 (50.55\%), 
    Swagger Core identifies seven out of 107 (6.5\%), and 
    Springdoc identifies 270 out of 277 (97.47\%).    
\end{itemize}

Respector can only generate OASs for the Java-based, Spring Boot, and Jersey framework-based APIs.
Swagger Core and Springdoc can only generate OASs for the Jersey and Spring Boot framework-based APIs, respectively. LRASGen can generate OASs for Java/Python/C\# APIs, and, through simple extension (creating the framework-specific criteria), other APIs can also be covered.

\begin{tcolorbox}[
    enhanced,
    width=\linewidth,
    colback=gray!10,
    colframe=gray!90,
    boxrule=1pt, 
    arc=4pt,
    boxsep=5pt
]
\textbf{Summary:}
LRASGen outperforms the three state-of-the-art OAS generation techniques, which,
on average, detect only
75.67\% endpoint methods, 75.24\% endpoint parameters, 4.78\% parameter constraints, and 26.75\% endpoint responses.

\end{tcolorbox}

\begin{table*}
    [!h] \footnotesize
    \centering
    \caption{Comparative results of LRASGen with Respector, Swagger Core, and Springdoc on 15 Java RESTful APIs}
    \label{TAB:ComparisonResults1}
    \setlength{\tabcolsep}{3.5mm}
    \begin{tabular}{llrr|rrrr}
        \hline
        \textbf{Objective}	&\textbf{API}	&\textbf{GT}	&\textbf{GT*}	&Respector	&Swagger Core	&Springdoc	&LRASGen \\	
        \hline								
\multirow{16}*{Endpoint Methods}	&Digdag	&41	&41	&\textbf{41}	&--	&--	&\textbf{41} \\	
	&enviroCar	&128	&128	&\textbf{128}	&--	&--	&\textbf{128}\\	
	&Features-Service	&18	&18	&\textbf{18}	&\textbf{18}	&NA	&\textbf{18} \\	
	&Gravitee	&28	&28	&\textbf{28}	&--	&--	&\textbf{28} \\	
	&Kafka	&74	&69	&\textbf{69}	&--	&--	&\textbf{69}\\	
	&Cassandra	&50	&50	&\textbf{50}	&--	&--	&\textbf{50} \\	
	&RESTCountries	&27	&27	&\textbf{27}	&22	&NA	&\textbf{28 (1)}\\	
	&Senzing	&34	&34	&\textbf{34}	&--	&NA	&\textbf{34} \\	
	&CatWatch	&14	&14	&\textbf{14}	&NA	&8	&\textbf{14} \\	
	&CWA	&5	&5	&\textbf{5}	&NA	&$\otimes$	&\textbf{5}\\	
	&OCVN	&278	&278	&\textbf{278}	&NA	&$\otimes$	&280 (4) \\	
	&Ohsome	&159	&159	&\textbf{159}	&NA	&\textbf{159}	&\textbf{159}\\	
	&Quartz	&14	&15	&14	&NA	&--	&\textbf{15} \\	
	&Proxyprint	&75	&75	&\textbf{75}	&NA	&--	&\textbf{77 (2)}\\	
	&Ur-Codebin	&7	&6	&\textbf{6}	&NA	&\textbf{6}	&\textbf{6}\\	\cline{2-8}
&\textbf{\textit{Sum}}		&952	&947	&\textbf{946}	&40	&173	&952 (7)	\\\hline
\multirow{16}*{Endpoint Parameters}	&Digdag	&96	&97	&80	&--	&--	&\textbf{97}\\	
        	&enviroCar	&351	&351	&202	&--	&--	&\textbf{351}\\	
        	&Features-Service	&35	&35	&\textbf{35}	&\textbf{35}	&NA	&\textbf{35}\\	
        	&Gravitee	&99	&99	&\textbf{99}	&--	&--	&\textbf{101 (2)}\\	
        	&Kafka	&175	&175	&154	&--	&--	&\textbf{175}\\	
        	&Cassandra	&94	&94	&\textbf{94}	&--	&--	&\textbf{94}\\	
        	&RESTCountries	&35	&37	&35	&34	&NA	&\textbf{37}\\	
        	&Senzing	&156	&160	&154	&--	&NA	&\textbf{160}\\	
        	&CatWatch	&32	&32	&\textbf{32}	&NA	&8	&\textbf{32}\\	
        	&CWA	&14	&14	&\textbf{13}	&NA	&$\otimes$	&\textbf{13}\\	
        	&OCVN	&5002	&5002	&\textbf{5002}	&NA	&$\otimes$	&4888\\	
        	&Ohsome	&1937	&1937	&\textbf{1914}	&NA	&\textbf{1914}	&1867\\	
        	&Quartz	&3	&7	&3	&NA	&--	&\textbf{7}\\	
        	&Proxyprint	&150	&168	&146	&NA	&--	&\textbf{161}\\	
        	&Ur-Codebin	&14	&14	&\textbf{14}	&NA	&14	&\textbf{14}\\	\cline{2-8}
&\textbf{\textit{Sum}}		&8193	&8222	&7977	&69	&1936	&\textbf{8032 (2)}	\\\hline
\multirow{16}*{Parameter Constraints}	&Digdag	&4	&112	&0	&--	&--	&\textbf{112}\\	
	&enviroCar	&12	&71	&0	&--	&--	&\textbf{71}\\	
	&Features-Service	&0	&37	&0	&0	&NA	&\textbf{37}\\	
	&Gravitee	&0	&59	&0	&--	&--	&\textbf{59}\\	
	&Kafka	&0	&265	&0	&--	&--	&\textbf{265}\\	
	&Cassandra	&3	&85	&0	&--	&--	&\textbf{85}\\	
	&RESTCountries	&12	&24	&12	&2	&NA	&\textbf{24}\\	
	&Senzing	&17	&87	&14	&--	&NA	&\textbf{87}\\	
	&CatWatch	&3	&9	&3	&NA	&1	&\textbf{9}\\	
	&CWA	&0	&11	&0	&NA	&$\otimes$	&\textbf{11}\\	
	&OCVN	&0	&4459	&0	&NA	&$\otimes$	&\textbf{4459}\\	
	&Ohsome	&58	&75	&0	&NA	&0	&\textbf{75}\\	
	&Quartz	&0	&228	&0	&NA	&--	&\textbf{228}\\	
	&Proxyprint	&0	&10	&0	&NA	&--	&\textbf{10}\\	
	&Ur-Codebin	&2	&14	&2	&NA	&0	&\textbf{14}\\	\cline{2-8}
&\textbf{\textit{Sum}}		&111	&5546	&31	&2	&1	&\textbf{5546}	\\\hline
\multirow{16}*{Endpoint Responses}	&Digdag	&41	&85	&41	&--	&--	&\textbf{85}\\	
	&enviroCar	&141	&201	&141	&--	&--	&\textbf{201}\\	
	&Features-Service	&18	&35	&18	&7	&NA	&\textbf{35}\\	
	&Gravitee	&26	&57	&26	&--	&--	&\textbf{57}\\	
	&Kafka	&108	&145	&69	&--	&--	&\textbf{145}\\	
	&Cassandra	&60	&87	&52	&--	&--	&\textbf{87}\\	
	&RESTCountries	&25	&72	&25	&0	&NA	&\textbf{72}\\	
	&Senzing	&34	&92	&34	&--	&NA	&\textbf{92}\\	
	&CatWatch	&10	&24	&10	&NA	&4	&\textbf{24}\\	
	&CWA	&7	&15	&5	&NA	&$\otimes$	&\textbf{15}\\	
	&OCVN	&278	&664	&278	&NA	&$\otimes$	&\textbf{664}\\	
	&Ohsome	&280	&449	&261	&NA	&261	&\textbf{449}\\	
	&Quartz	&75	&127	&75	&NA	&--	&\textbf{127}\\	
	&Proxyprint	&14	&22	&14	&NA	&--	&\textbf{22}\\	
	&Ur-Codebin	&7	&12	&6	&NA	&5	&\textbf{12}\\	\cline{2-8}
&\textbf{\textit{Sum}}		&1124	&2087	&1055	&7	&270	&\textbf{2087}	\\\hline

    \end{tabular}
\end{table*}

\begin{table*}
    [!t] \footnotesize
    \centering
    \caption{Comparative results of LRASGen with Respector, Swagger Core, and Springdoc on four Python and one C\# RESTful APIs}
    \label{TAB:ComparisonResults2}
    \setlength{\tabcolsep}{4.2mm}
    \begin{tabular}{llrr|rrrr}
        \hline
\textbf{API}	&\textbf{Objective}	&\textbf{GT}	&\textbf{GT*}	&Respector	&Swagger Core	&Springdoc	&LRASGen \\		
        \hline									
\multirow{6}*{Endpoint Methods}	        &Poke	&NA	&97	&NA	&NA	&NA	&\textbf{97}	\\	
	        &Gramps	&NA	&148	&NA	&NA	&NA	&\textbf{148}	\\	
	       &Jupyter	&NA	&28	&NA	&NA	&NA	&\textbf{28}	\\	
	        &Mlmmj	&NA	&9	&NA	&NA	&NA	&\textbf{9}	\\	
	        &Bitwarden	&NA	&74	&NA	&NA	&NA	&\textbf{74}	\\	\cline{2-8}
	&\textbf{\textit{Sum}}	&NA	&356	&NA	&NA	&NA	&\textbf{356}	\\	        \hline
\multirow{6}*{Endpoint Parameters}	        &Poke	&NA	&192	&NA	&NA	&NA	&\textbf{192}	\\	
       	        &Gramps	&NA	&488	&NA	&NA	&NA	&\textbf{488}	\\	
       	       &Jupyter	&NA	&20	&NA	&NA	&NA	&\textbf{20}	\\	
       	        &Mlmmj	&NA	&54	&NA	&NA	&NA	&\textbf{54}	\\	
       	        &Bitwarden	&NA	&86	&NA	&NA	&NA	&\textbf{86}	\\	\cline{2-8}
	&\textbf{\textit{Sum}}	&NA	&840	&NA	&NA	&NA	&\textbf{840}	\\	        \hline
\multirow{6}*{Parameter Constraints}	        &Poke	&NA	&49	&NA	&NA	&NA	&\textbf{49}	\\	
       	        &Gramps	&NA	&84	&NA	&NA	&NA	&\textbf{84}	\\	
       	       &Jupyter	&NA	&2	&NA	&NA	&NA	&\textbf{2}	\\	
       	        &Mlmmj	&NA	&1	&NA	&NA	&NA	&\textbf{1}	\\	
       	        &Bitwarden	&NA	&60	&NA	&NA	&NA	&\textbf{60}	\\	\cline{2-8}
	&\textbf{\textit{Sum}}	&NA	&196	&NA	&NA	&NA	&\textbf{196}	\\	        \hline
\multirow{6}*{Endpoint Responses}	        &Poke	&NA	&97	&NA	&NA	&NA	&\textbf{97}	\\	
        	        &Gramps	&NA	&591	&NA	&NA	&NA	&\textbf{591}	\\	
        	       &Jupyter	&NA	&56	&NA	&NA	&NA	&\textbf{56}	\\	
        	        &Mlmmj	&NA	&9	&NA	&NA	&NA	&\textbf{9}	\\	
        	        &Bitwarden	&NA	&247	&NA	&NA	&NA	&\textbf{247}	\\	\cline{2-8}
	&\textbf{\textit{Sum}}	&NA	&1000	&NA	&NA	&NA	&\textbf{1000}	\\	        \hline

    \end{tabular}
\end{table*}

\subsection{Threats To Validity}

This section discusses some possible threats to the validity of our work, and how these were mitigated.

\subsubsection{\textbf{Internal Validity}}

The internal validity is crucial for assessing the precision and reliability of the results:
In this study, the internal validity concerns the consistency and accuracy of the LRASGen-generated API specifications, and the extent to which the findings can be trusted as a true reflection of the effectiveness.

The LRASGen precision when detecting parameter constraints and endpoint responses was high, with an average of 100\%, for both.
However, the precision when identifying endpoint methods and endpoint parameters was comparatively lower, with 99.46\% for both GPT-4o mini and DeepSeek V3.
This suggests that LLMs may have varying levels of proficiency regarding different aspects of API specifications, which could be a threat to internal validity.
To address this, the study examined the performance of two different LLMs, providing a more robust assessment of the reliability of the LRASGen approach.

The study also considered the threat posed by the dynamic nature of software development, where RESTful APIs are frequently updated and evolved.
LRASGen's ability to accurately generate specifications from source code depends on the completeness and accuracy of the source code:
To ensure the validity of the results, the most recent implementations of the RESTful APIs were used.

Finally, internal validity also depends on the accuracy of the ground truth used for comparison.
The ground truth for each API was established through analysis of both the developer-provided documentation and the source code.

\subsubsection{\textbf{External Validity}}

The external validity is essential for the generalizability of the results, and the model's applicability across diverse real-world contexts:
It concerns the extent to which the results can be generalized beyond the specific conditions under which the research was conducted.

One potential threat to external validity relates to the diversity and variability of programming languages, frameworks, and development practices:
LLMs trained on certain types of code, or within specific domains, may not perform as effectively when applied to different code or domains.
This variability could limit the model's effectiveness in a broader range of development environments.

Another threat relates to the quality and representativeness of the training data:
LLMs are only as good as the data that they are trained on.
If the training data is biased towards certain types of code, languages, or structures, then the model may not generalize well to other types.
Ensuring a broad and representative training dataset is essential to address this concern.

The dynamic nature of software development also poses a threat to external validity.
The field is constantly evolving, with new languages, tools, and best practices emerging frequently:
LLMs must be able to adapt to these changes to maintain relevance and effectiveness.

Finally, the specific context in which the LLM is deployed can also influence its performance.
Factors such as the development team's expertise, the complexity of the project, and the existing development processes can all impact on how well the LLM's assistance is integrated and used:
This contextual variability may affect the model's ability to provide consistent assistance across different teams or projects.

\section{Discussion
\label{sec:7}}

This section discusses the three main reasons why LRASGen generated imprecise specifications.
The first one, during the dataset-preparation phase, was caused by the source code not being properly organized.
Taking the Java Jersey API as an example:
\textit{import} was used as the entry point to traverse and obtain all the source code, which initially resulted in the LLM not having enough information to analyze the code content.

The prompt-writing and parameter settings also led to instability in what the LLM returned:
The initial prompt was very simple, for example: 
\textit{``Please tell me what endpoints are in the code?''}
Also, the \textit{temperature} parameter value was set to 1.2, which led to inconsistent results each time (too many, too few, and non-compliant results).
Finally, overly loose and unclear requirements also led to excessive output:
Initially, the LLM was asked to answer questions like:
\textit{``What are the parameters of a certain endpoint and what are the constraints on the parameters?''}
However, the word ``constraints'', as a technical term redefined in our task, was not fully understood by the LLM, resulting in the outputs not fully satisfying our requirements.
Through systematic experimentation with parameter tuning, it was determined that when the temperature parameter was set to 0.2, the LLM could consider both accuracy (answering accurately) and correctness (verified correctness).
According to our experimental results, false positives (seven of the 1308 endpoint methods, accounting for 0.53\%; and two of the 8872 endpoint parameters) are extremely rare.


It was found that LRASGen had generated many more parameter constraints and method responses than the ground truth:
The developers examined these constraints and responses in the source code, and reported that:
1) Most of the constraints were for the code (business logic), database settings (such as column length), and enumeration; and 
2) Most of the missing responses had response status codes \texttt{4xx} (client-error responses) or \texttt{5xx} (server-error responses).
When writing specifications, some developers may just focus on response status code \texttt{200} (successful responses), ignoring response status codes \texttt{4xx} and \texttt{5xx}.

\section{Related Work
\label{sec:8}}

\subsection{Respector}

Respector \cite{huangGeneratingRESTAPI2024} is an innovative tool designed to automatically generate RESTful API specifications through static analysis.
Respector uses static program analysis, examining the source code of RESTful APIs to infer their specifications.
It is capable of identifying endpoint methods, parameters, responses, and even the conditions under which these endpoints return successful or erroneous HTTP responses.
It uses both static and symbolic analysis to extract metadata from the API's, generating an OpenAPI specification that accurately reflects the API's functionality.

A key strength of Respector is its ability to handle real-world APIs with complex structures and dependencies.
It can identify elements with a high degree of precision and recall.
However, situations where Respector may fail to produce accurate OpenAPI specifications include:
When the class file lacks sufficient annotations;
when parameters and responses are not explicitly defined; or 
when the code structure is excessively complex and nested beyond the Respector's pre-trained knowledge boundaries. 

\subsection{AutoOAS}

AutoOAS \cite{lercherGeneratingAccurateOpenAPI2024} is another static-analysis approach, for automatically generating OpenAPI specifications from Java Spring Boot projects.
Other approaches, such as Respector \cite{huangGeneratingRESTAPI2024}, Prophet \cite{cernyStaticCodeAnalysis2024}, and Springdoc \cite{springdocopenapiSpringdocOpenAPI}, may ignore Spring profiles, misrepresent data models, or fail to handle exception-driven HTTP response codes.
However, AutoOAS uses the Spoon framework to parse Java source code, detect endpoints, parameters, and responses, and generate OpenAPI descriptions.

AutoOAS addresses gaps in automated API documentation, particularly for projects using Spring profiles and complex data models, 
while maintaining reasonable runtime performance (median 7.6 seconds).
It advances the state-of-the-art in RESTful API-specification generation by combining static analysis precision with comprehensive support for modern Spring Boot features.
However, its shortcomings include: 
It is deeply integrated with the Spring Boot framework and cannot work across other languages and frameworks.

\subsection{NLPtoREST}

NLPtoREST \cite{kimEnhancingRESTAPI2023} is an innovative method to improve the testing of RESTful APIs, using natural language processing (NLP) to enhance the OpenAPI specifications.
It begins by extracting additional rules from the human-readable part of the OpenAPI specifications, which is something often overlooked by other testing tools \cite{arcuriAutomatedBlackWhiteBox2021,corradiniAutomatedBlackboxTesting2023,atlidakisRestlerStatefulRest2019}.
NLPtoREST adds these extracted rules, such as parameter constraints and inter-parameter dependencies, to the original machine-readable specification to help test generators create more accurate test cases.

The approach consists of two major steps: 
rule extraction and rule validation.
In the extraction step, NLPtoREST analyzes the natural language descriptions in the OpenAPI specifications using a custom-trained Word2Vec model:
It identifies keywords, parameter names, values, and inter-parameter dependencies.
This is followed by the validation step, which addresses any inaccuracies from the rule extraction.
The validation process removes spurious or incorrect rules by executing test cases against a deployed API instance, and pruning incompatible rules through static and dynamic analysis.

A key innovation is its ability to integrate natural language descriptions into the test-generation process.
Unlike other methods like RestCT (which relies on rigid pattern matching), NLPtoREST applies a flexible, semantic approach (using NLP techniques) to infer parameter rules and dependencies.
NLPtoREST has proven to be significantly more effective in extracting inter-parameter dependencies, a critical aspect of accurate test generation.

NLPtoREST was empirically tested on various real-world APIs, with the results demonstrating substantial improvements in test-generation performance.
Using the NLPtoREST-enhanced specifications resulted in higher code coverage, reduced numbers of invalid requests, and more unique faults being found.
This suggests that NLPtoREST can significantly enhance the effectiveness of automated RESTful API testing.

\section{Conclusions and Future Work
\label{sec:9}}

In this paper, we have introduced LRASGen, a novel approach that generates OpenAPI specifications using LLMs.
Using a framework-knowledge database, LRASGen extracts and cleans endpoint code files, and then submits these codes as context to the LLM.
Compared with previous studies \cite{huangGeneratingRESTAPI2024,swaggerSwagger,springdocopenapiSpringdocOpenAPI}, LRASGen only needs RESTful API source code as input, without a need for the runtime environment, or to fine-tune the LLM.
It can also easily be extended to different programming languages and frameworks (only one framework-specific criterion needs to be established, which can also be done with the help of the LLM).

We reported on an analysis of the OpenAPI specification-generation performance of LRASGen, with GPT-4o mini or DeepSeek V3.
Based on the experimental results, it was observed that, for all 20 APIs, the LRASGen-generated specification included all the developer-provided specifications, and were more detailed (especially for the parameter constraints and endpoint responses);
for the 15 Java RESTful APIs analyzed, LRASGen demonstrated consistent superiority across all four identification tasks (endpoint methods, endpoint parameters, parameter constraints, and endpoint responses),
significantly outperforming Respector, Swagger Core, and Springdoc.
LRASGen also demonstrated cross-language capability compared with the baseline tools (Respector, Swagger Core, Springdoc), which failed to process the four Python and one C\# APIs due to language and framework constraints:
LRASGen consistently generated OAS specifications in these situations.

Some remaining future work for LRASGen includes:
The analysis and cleaning of the source code aims to avoid problems such as token-limitation issues, attention-distraction, and information-forgetting.
However, these issues may persist (and even escalate) as the source code expands in size and complexity.
In future work, we can further study how to simplify and refine the source code to avoid these problems.
We also found that the overall running time required (during the experiments) was related to the LLM's computing capacity, source code's size and complexity.
Future work can further examine how to adjust the LLM's structure to improve the computing capabilities, and how to simplify and refine the source code to reduce running time.

\section*{Acknowledgment}
This work was partly supported by the Science and Technology Development Fund of Macau, Macau SAR, under Grant No. 0021/2023/RIA1.

\bibliographystyle{ACM-Reference-Format}
\bibliography{reference}

\end{document}